\documentclass{WileyChemistry-template}
\usepackage{cleveref}
\usepackage{xr}
\title{\raggedright Catalight -- an open source automated photocatalytic reactor package illustrated through plasmonic acetylene hydrogenation}

\author{
\begin{minipage}{\textwidth}
    B. B. Bourgeois,*\textsuperscript{[a]}\textsuperscript{[$\alpha$]} 
    A. X. Dai,\textsuperscript{[a]} 
    C. C. Carlin,\textsuperscript{[a]} 
    L. Yuan,\textsuperscript{[a]} 
    A. Al-Zubeidi, \textsuperscript{[a]}\textsuperscript{[$\beta$]} 
    W-H. Cheng,\textsuperscript{[b]} 
    D. F. Swearer,\textsuperscript{[c]} 
    J. A. Dionne*\textsuperscript{[a]} 
\end{minipage}
}

\newcommand{\affiliation}{
\begin{itemize}

\item[{[a]}] B. Bourgeois*, A. Dai, C. Carlin, L. Yuan, A. Al-Zubeidi, Prof. J. Dionne*\\
Department of Materials Science and Engineering, Stanford University
450 Jane Stanford Way, Stanford, CA 94305
E-mail: bbourge6@stanford.edu; jdionne@stanford.edu

\item[{[b]}] Prof. W.-H. Cheng\\
Department of Materials Science and Engineering, National Cheng Kung University\\
No. 1, Dasyue Rd, East District, Tainan City, Taiwan 701

\item[{[c]}] Prof. D. Swearer\\
Department of Chemical and Biological Engineering, Northwestern University\\
633 Clark St, Evanston, IL 60208

\item[{[$\alpha$]}] Current Affiliation: National Renewable Energy Laboratory, Golden, CO

\item[{[$\beta$]}] Current Affiliation: ExxonMobil technology and Engineering Company, Annandale, NJ
\end{itemize}
}

\renewcommand{\dedication}{
	\begin{minipage}{\textwidth}
	\end{minipage}
}

\renewcommand{\abstract}{An open-source and modular Python package, Catalight, is developed and demonstrated to automate (photo)catalysis measurements. (Photo)catalysis experiments require studying several parameters to evaluate performance, including temperature, gas flow rate and composition, illumination power, and spectral profile. Catalight orchestrates measurements over this complicated parameter space and systematically stores, analyzes, and visualizes the resulting data. To showcase the capabilities of Catalight, we perform an automated apparent activation barrier measurement of acetylene hydrogenation over a plasmonic AuPd catalyst on Al$_{2}$O$_{3}$ support, simultaneously varying laser power, wavelength, and temperature in a multi-day experiment controlled by a simple Python script. Our chemical results unexpectedly show an increased activation barrier upon light excitation, contrary to previous findings for other plasmonic reactions and catalysts. We show that the reaction rate order with respect to both acetylene and hydrogen is unchanged upon illumination, suggesting that molecular surface coverage is not changing under light excitation. By analyzing the inhomogeneity of the laser induced heating, we attribute these results to a partial photothermal effect combined with a photochemical/hot electron driven mechanism. Our findings highlight the capabilities of a new experiment automation tool; explore the photocatalytic mechanism for an industrially relevant reaction; and identify systematic sources of error in canon photocatalysis experimental procedures.
}

\newcommand{\keywords}{
    \textbullet\ Automation 
    \textbullet\ Bimetallics
    \textbullet\ Heterogeneous catalysis 
    \textbullet\ Plasmonic photocatalysis
    \textbullet\ Python 
}

\begin{document}

\twocolumn[\vspace{-1.5cm}\maketitle\vspace{-1cm}
	\textit{\dedication}\vspace{0.4cm}]
\small{\begin{shaded}
		\noindent\abstract
	\end{shaded}
}

\begin{figure} [!b]
\begin{minipage}[t]{\columnwidth}{\rule{\columnwidth}{1pt}\footnotesize{\textsf{\affiliation}}}\end{minipage}
\end{figure}




\section*{Introduction}
\label{introduction}

Catalysis serves a critical role in the development of the modern world, providing control over the speed and selectivity of chemical reactions. However, due to the expansive parameter space of catalysis, the cost of discovering and optimizing new catalytic materials, in both work-hours and dollars, has not been small. Most famously, Alwin Mittasch performed nearly 6,500 experiments on 2,500 materials systems to discover the catalyst used in the industrial Haber-Bosch ammonia synthesis process.\cite{farberChemistryPhilosophyWay1966} Over 100 years later, scientists and engineers are attempting to integrate advances in artificial intelligence and computer control systems to automate materials development, testing, and analysis, removing human time demands from catalyst development.\cite{kumarCatalysisinaBoxRoboticScreening2020, burgerMobileRoboticChemist2020, szymanskiAutonomousLaboratoryAccelerated2023, koscherAutonomousMultipropertydrivenMolecular2023, koderaAutomaticRobotSystem2023, Kuijpers2018} In many cases, sophisticated, purpose-built, self-driving laboratories utilize expensive robotics and advanced computational methods. While pursuing fully autonomous laboratories promises a future paradigm shift in materials research, modern software tools and computer-interfaced hardware greatly simplify the prospects of democratizing automation, even in traditional experimental laboratories.\cite{Christopher2020, ozinHomoDeusAm2023, wangDesigningMaterialsAcceleration2023}

Photochemistry specifically is positioned to benefit from experimental automation. A huge technical challenge for researchers performing this work is the vast experimental landscape associated with photocatalysis. Light power, photon flux, and wavelength; catalyst temperature; gas flow rate; and reactant composition all have significant impacts on the behavior of a catalytic system, necessitating considerable investment of time and resources to explore a single catlyst or reaction. Automating such experimentation opens the door for integrated machine learning models to speed up the development of increasingly complex catalytic systems.\cite{Zhao2023, Li2018, MOSES2023, Yang2020} Furthermore, photochemistry experiments, especially in heterogeneous catalysis, can require custom hardware configurations which do not natively support automation or even software integration. As a result, researchers often manually adjust hardware settings with significant periods of downtime while waiting for reaction conditions to reach a steady state. 

Automation not only saves countless researcher-hours, but it also provides precision reproducibility, removing the potential for human error. Some important experimental procedures may even be prohibited by difficult experimental conditions without automation protocols. Kim \textit{et al.} write of the potential importance of “apparent activation energy” measurements in plasmonic photocatalysis experiments.\cite{Kim2016} In this procedure, a given light intensity and wavelength is used to illuminate the catalyst sample while the temperature is modulated. The change in the chemical reaction rate as a function of the change in temperature provides the activation energy. The activation barrier also appears to change with the illumination conditions, suggesting that photons may fundamentally impact the reaction mechanism. In seminal work by Zhou \textit{et al.}, apparent activation barrier measurments were repeated 46 times, equating to 184 measurements under different optical conditions and temperatures.\cite{Zhou2018} Utilizing a gas chromatograph (> 10 min sampling time), a reasonable steady state time (30 min), and 3-4 replicate data points per experimental condition, generating a single data set can take well over 200 continuous hours of lab time, most of which is waiting for steady state conditions. Scaling this type of experimental procedure is not feasible. Furthermore, while some hardware components provide basic automation capabilities through the manufacturer provided software, it is very uncommon to support linking automated procedures between multiple disparate pieces of equipment. Researchers end up working between multiple software windows to independently control heating systems, flow controllers, and light sources. To address this challenge, we developed Catalight, an open-source Python package that integrates common photochemistry hardware and orchestrates automated catalysis experiments and subsequent data analysis.\cite{catalight_package}

The focus of this work is to highlight the main feature of Catalight, and deploy this software for an automated experimental demonstration. A comprehensive guide to utilizing the tool is provided in the online documentation. \cite{catalight_rtd} We study the hydrogenation of acetylene ($C_2H_2$) into ethylene ($C_2H_4$) over a plasmonic AuPd alloy catalyst. Selective hydrogenation of acetylene is simultaneously an important industrial chemistry and an academic model system.\cite{McCue2015, takhtravanchiAcetyleneSelectiveHydrogenation2018} Au, Ag, and Cu alloyed with Pd are frequently studied and bimetallic AgPd catalysts are commercially used to thermally drive this reaction.\cite{zhangSelectiveHydrogenationAcetylene2014, ballAgPdCuPdCatalysts2020, peiAgAlloyedPd2015a} Coincidentally, these alloys sustain localized surface plasmon resonances due to the significant coinage metal content. Reports of plasmonic selectivity enhancement for acetylene hydrogenation chemistry on antenna-reactor style plasmonic catalysts naturally leads to a question of whether these optimally performing thermal catalysis alloys could potentially be further improved through use of their inherent plasmonic properties.\cite{sytwuBimetallicNanostructuresCombining2019, swearerHeterometallicAntennaReactor2016} In this work, we thoroughly investigate the performance of 6\% Pd on Au dilute alloy nanoparticles on $\gamma Al_2O_3$ support using the Catalight system. Contrary to results found in other plasmon-activated catalysts, we found evidence of plasmon-enhanced photothermal chemistry and we did not observe greater photochemical selectivity toward ethylene for this AuPd catalyst under the parameters explored in this study.

\section*{Results and Discussion}
\label{results_discussion}

This report is split into two distinct sections. The first section describes the design philosophy of the Catalight package while the second section describes experimental results exploring the photocatalytic hydrogenation of acetylene over a plasmonic AuPd catalyst. 

\subsection*{Catalight Design Philosophy}
\label{hardware_control}

Catalight is a Python software package designed with modularity and approachability in mind. The package is split into two main components: hardware control and data analysis. This split structure is found in both the structure of the code, split into catalight.equipment and catalight.analysis subpackages, and the usage of the software. Users are intended to first design and actuate measurements, then separately run analysis on the collected data. Because the data output by Catalight could vary for different hardware configurations, keeping the experimentation and analysis separate maximizes the flexibility of the code. 

The approach we've taken to hardware control plays into the advantages of using Python as a programming language.\cite{guoTenMillionUsers2021, ayerWhyScientistsShould2014, stewartPythonScientists2023} Python has remained one of the world's most popular programming languages, particularly amongst scientists and engineers. With that accolade comes a huge community of open-source packages for scientific computing and hardware interfacing.\cite{closNMRbotPythonScripts2013, starkeyScriptedControlSystem2013, binderQudiModularPython2017, johnsonPLACEOpenSourcePython2015} Additionally, many universities offer or require Python-based programming courses for undergraduate STEM majors. Catalight is designed to leverage the existing tools to the greatest degree possible. The primary design goal of Catalight is to provide a basic architecture for running arbitrary catalysis experiments in a structured way. 

\begin{figure*}
\begin{center}
\includegraphics[width=17.4cm]{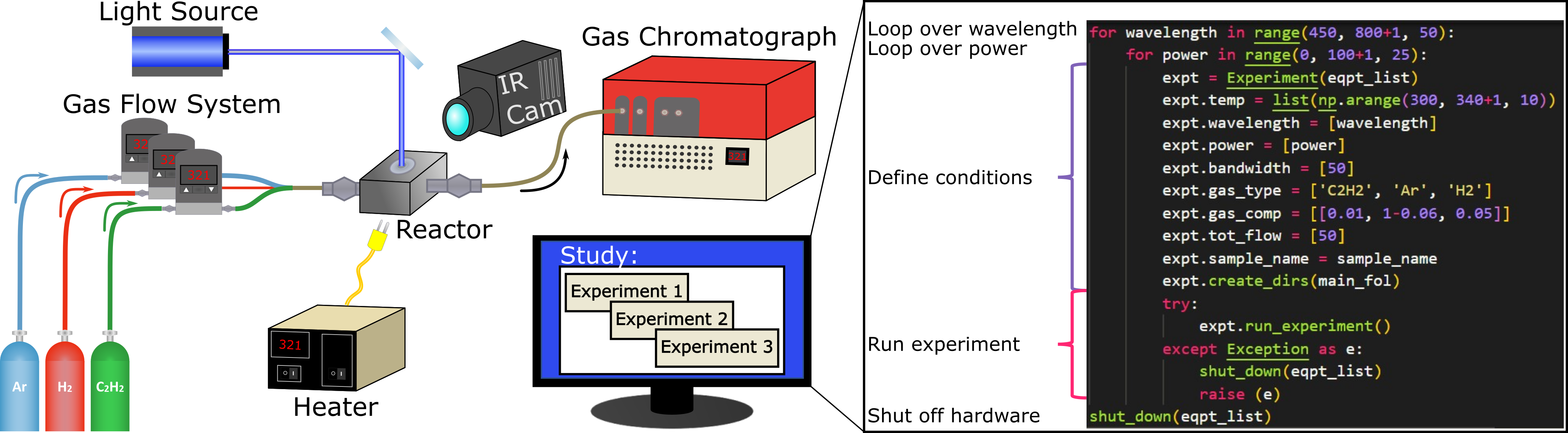}
\caption{A conceptual overview of the Catalight automation system. Virtual “experiments” are created defining a user’s desired parameter sweep, and Python connections of physical hardware are passed to each experiment. Experiment execution is handled programmatically for an arbitrary number of experiments and the uniform data collection style allows for automated data analysis.}
\label{fig1}
\end{center}
\end{figure*}

This concept is achieved by separating the performance of an experiment out from the control of an individual piece of equipment. Each physical piece of equipment is mimicked as a virtual Python object, which is instantiated within the software to serve as the controller for the physical instrument. Experiments are then created as a separate Python class. These “experiment” objects utilize existing equipment objects and represent each parameter sweep the user would like to perform. The user provides instructions for how the experiment should be performed by editing the experiment object attributes, either through scripting or utilizing the Catalight graphical user interface (GUI). When the experiment is run (by calling its “run\_experiment” method), the appropriate instructions for carrying out the experiment are passed to the connected hardware. The user is able to generate an arbitrary number of experiments, a grouping of which we generally refer to in the documentation as a “study”. \cref{fig1} shows a graphical outline of this process applied to the lab space used in this work. The included code is taken from the script used for the measurements made in the apparent activation barrier measurements section. Virtual experiments are created in a loop through various laser powers and wavelengths. For each set of optical conditions, a temperature sweep experiment is created to probe the activation barrier of the system.

The initial goal for the Catalight project was to automate the hardware specifically used in our laboratory (and illustrated in \cref{fig1}), but this was achieved with future expansion of the package in mind. The power of object-oriented programming is that many tools can be re-utilized. The goal here is to provide a toolbox which can help other labs utilize the transferable parts of the package we have already developed while writing their own plug-ins for their specific hardware. To this end, early adopters of the software will need to spend some time writing their own code to integrate with the rest of Catalight. A large motivation for choosing to build this system with Python is that a number of hardware providers already provide some applications programming interfaces (APIs) for their equipment which can be accessed via Python. 

To date, the package currently supports a number of specific hardware devices including Alicat mass flow controllers, Thorlabs diode laser drivers interfaced via an MCC DAQ board for communication, NKT Photonics lasers, Watlow heating systems, and SRI gas chromatographs. While these specific tools are commonly used in the plasmonic photocatalysis community, these pieces of hardware can be abstracted to fit into the rest of the codebase without consideration for specific branding. For example, an MKS branded flow controller could be interacted with throughout the code in the exact same manner as an Alicat brand flow controller, provided their control code is written in a uniform format. This type of generalization should greatly simplify the implementation of abstract flow controllers, light sources, analysis devices, and heating systems that can all interact with the central "experiment" class in an identical manner.

\begin{figure}
\begin{center}
\includegraphics[width=8.6cm]{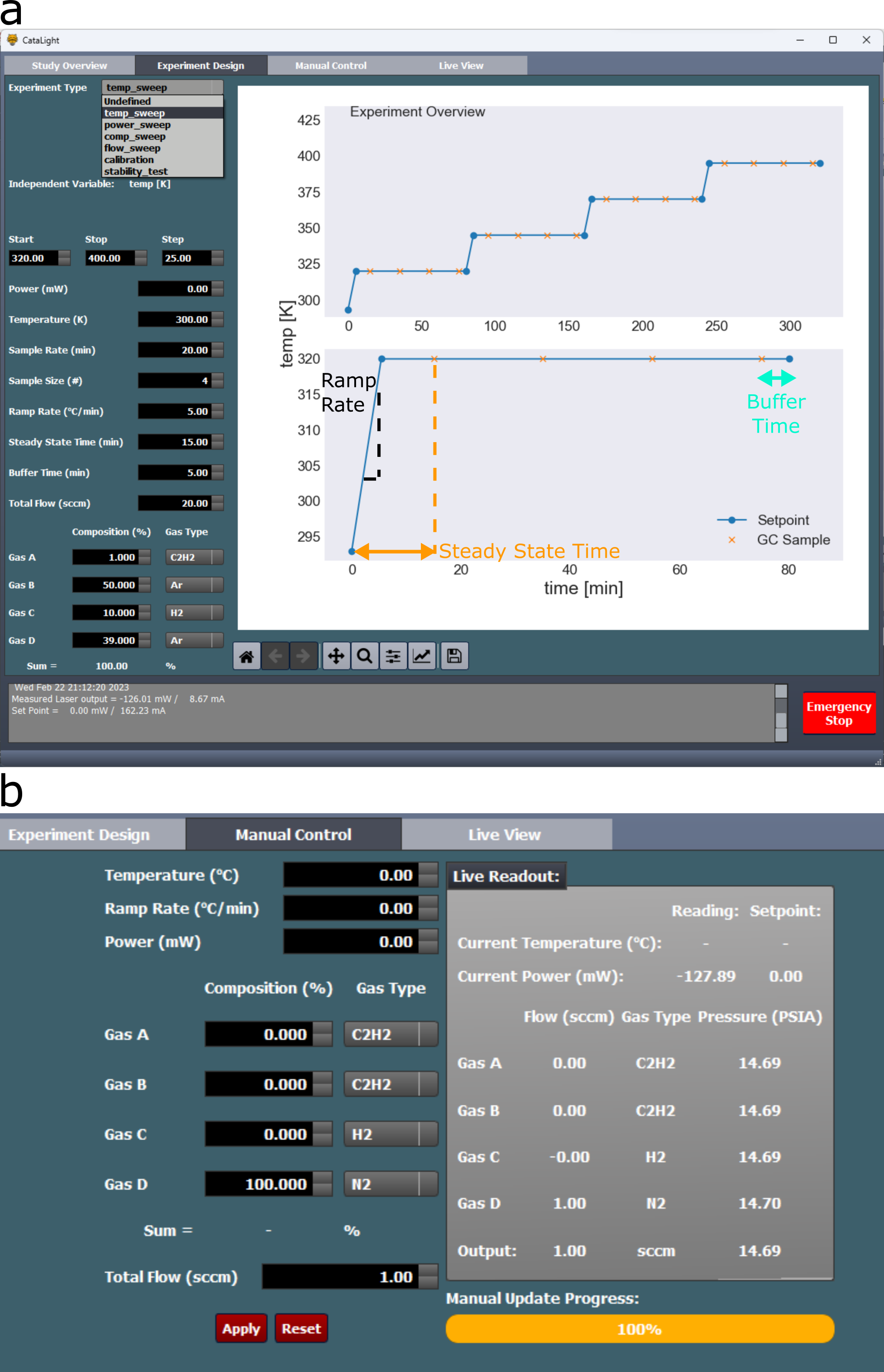}
\caption{Screenshots of the Catalight Graphical User Interface (GUI). A)  “Experiment Design” tab used to formulate an individual experiment. B) “Manual Control” tab used to interact with the hardware in real time.}
\label{fig2}
\end{center}
\end{figure} 

The Catalight package includes a GUI to make using the system more approachable for researchers who do not have Python experience. \cref{fig2} provides compiled screenshots of the main Catalight GUI used for the design and implementation of experiments. This figure is also illustrative of how interaction with the software would work in a scripted version. The user creates experiment objects, compiled into a list on the “Study Overview” tab. The experiment parameters are then defined in the “Experiment Design” tab seen in \cref{fig2}a. The user selects from a list of experiment types which generally consists of parametric studies. The fillable form on the left-hand side of the GUI updates to reflect the selection, and the user enters both the constant and variable parameters for the chosen experiment. In a scripted format, this would be achieved by defining the same parameters by directly assigning values to attributes of an experiment object. A preview of the experimental procedure is plotted on the right-hand side of the GUI. A major advantage of the GUI is the ability to control equipment in real time with active status feedback, as demonstrated in \cref{fig2}b which depicts the “Manual Control” tab.


In addition to saving hours of hands-on experimental time, Catalight, and the automation of experimental procedures in general, allows for systematic data handling and processing. We have elected to keep data collection in its native format for each device. Instead of converting data into a single uniform data type, the Catalight system directs the output of the connected devices to specific folders for each step of an experiment. This data-type agnostic approach maximizes the flexibility of the codebase by simply requiring the users to point any new machines to the active data collection path. The user can then write a new code addition to support the new data types, which would be hard to envision at the onset of such a development process. For example, we have written a number of analysis tools for gas chromatography (GC) data--which consists of intensity vs. time data sets--including a relatively complex calibration process which is only applicable to GC-type data. Future users may be interested in \textit{in situ} Fourier-Transform Infrared (FTIR) spectroscopy, for example, which is not obviously connected to GC data in format. The logic here is to save the relevant FTIR spectra in the active folder during collection rather than developing an overly complex analysis script that treats both data types with some equivalency. A user can write a custom script that analyzes data during an experiment as part of a feedback system, but such functions were not explicitly designed for in this version of the software.

\begin{figure*}
\begin{center}
\includegraphics[width=12cm]{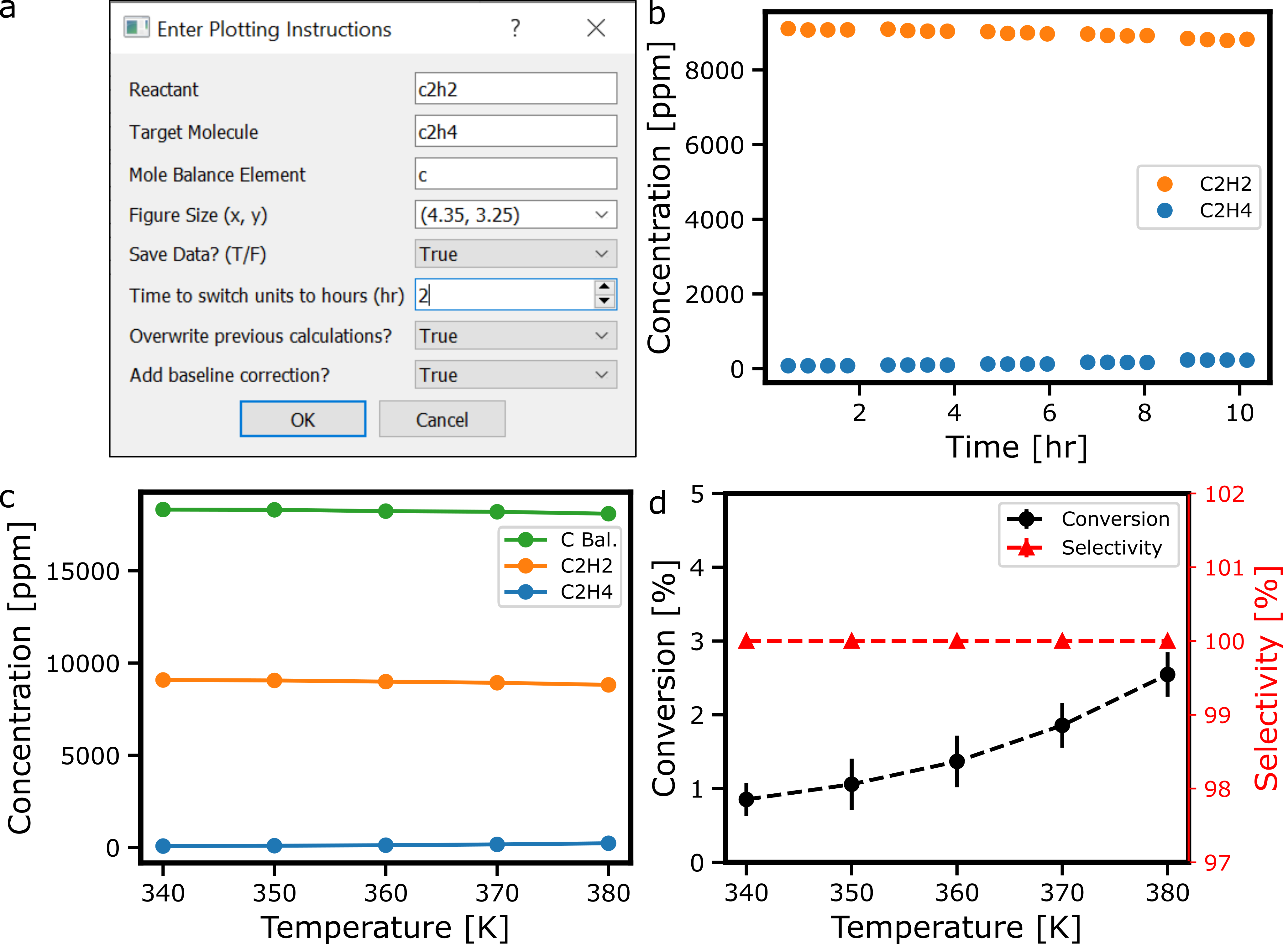} 
\caption{An example of the analysis of a single activation barrier experiment within Catalight. A) A simple dialog window used to provide analysis instruction B) Raw concentrations from GC samples and C) concentrations as a function of independent variable collected during an automated temperature sweep. The green line depicts the total mole balance element selected - carbon in this case. D) Calculated conversion and selectivity of the reactant (C$_2$H$_2$) into the target molecule (C$_2$H$_4$).}
\label{fig3}
\end{center}
\end{figure*}

Because data collected via Catalight is stored in a structured and consistent format, automated analysis is a straightforward process. Basic analysis procedures have already been implemented within Catalight for GC data collected via an FID detector on an SRI-brand machine. \cref{fig3} depicts the initial analysis procedure for such data and shows a subset of the data collected in the activation barrier experiment described shortly here after.  A simple dialog is used to request a path to the user’s data folder and calibration file, then the user fills in the form shown in \cref{fig3}A. The plots in \cref{fig3}B-D are automatically generated and saved in the "Results" folder for every experiment folder located in the directory selected by the user. As shown in \cref{fig3}B, the gas chromatograms are converted into concentrations for each detected gas species based on the calibration file provided and plotted vs. total experiment runtime. In this case, only acetylene and ethylene are detected, though ethane is also observed as a byproduct. No C3 or C4 species are detected. \cref{fig3}C computes the average concentration of each species as a function of the experimental step. This data depicts a temperature sweep experiment. The unlabeled, green line depicts the chosen mole balance element - a carbon balance in this case. The carbon concentration is automatically determined using a regular expressions interpretation of each detected molecule. The code automatically assesses the number of carbon atoms in the molecular formula. Finally, in \cref{fig3}D the conversion and selectivity of the reaction is determined based on the specified reactant and target molecules. This experiment shows 100\% chemical selectivity for the conversion of acetylene into ethylene and an exponential increase in the conversion with increased temperature. The subfigures displayed in \cref{fig3} are shown exactly as automatically processed by the code, after running an additional low conversion rescaling script (also included in the package). Additional scripts are provided within the Catalight package to perform routine analyses such as plotting raw chromatograms, plotting multiple experiments in a single figure, and relabeling the x-axis data. In total, the Catalight package provides a comprehensive toolkit for orchestrating physical experiments, saving and organizing the produced data, and performing basic catalytic data analysis. In the current form,this toolbox is directly applicable to labs that possess the same types and brands of equipment as ours. However, we have designed the package with adaptability in mind to minimize the barrier to access for labs that wish to incorporate additional equipment or devices from other vendors. The structure of the system should remain the same, while specific interfaces need to be developed to expand the capabilities. With some additional effort, such a system could be employed to realize adaptive experiments, in which experimental parameters are chosen in live feedback with the catalytic results. Machine learning methods could be implemented as a method for rapidly exploring and optimizing across a vast parameter space.

\subsection*{Apparent Activation Barrier Measurements} \label{experimental}
\begin{figure}
\begin{center}
\includegraphics[width=8.6cm]{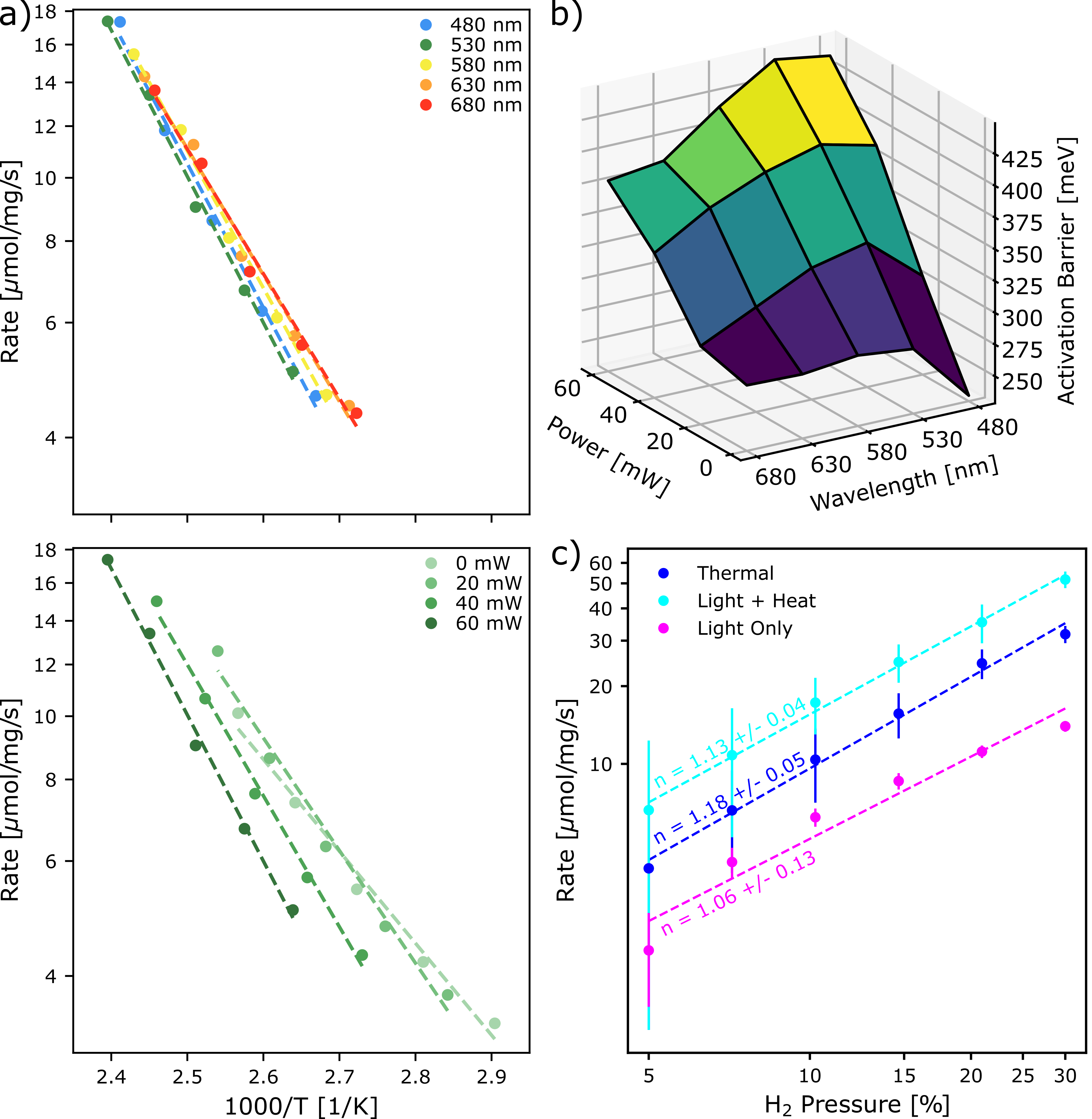}
\caption{A) Representative Arrhenius fits for 60 mW at various central wavelengths and a center wavelength of 530 nm for various laser powers. B) Resulting apparent activation barrier as a function of power and wavelength C) H$_2$ rate order for a thermal reaction at 340 K, a photoreaction at 300 K, and a combined reaction at 340 K under illumination.}
\label{fig4}
\end{center}
\end{figure}

To demonstrate the capabilities of the Catalight system, we performed automated testing of a plasmonic $\approx{}$5 nm Au$_{94}$Pd$_6$ catalyst. The catalyst was synthesized using colloidally grown nanoparticles using the method described by Wrasman et al. \cite{wrasmanSynthesisColloidalPd2018a, sunfacilesynthesismonodisperse, cargnelloSynthesisColloidalPd}. After synthesizing and washing, the nanoparticles were loaded onto a nanostructured $\gamma Al_2O_3$ support powder by wet impregnation at a ratio of 4 wt\% catalyst. Full details are documented in the supplementary information section. The dried catalyst is heat treated at 300 \textdegree{C} for 30 min in air to remove remaining surfactant. 2.5 mg of this catalyst powder is loaded into a Harrick in-situ reactor with a domed lid. An NKT FIU-15 supercontinuum laser filtered by a variable bandpass filter is incident on the sample through a transparent SiO$_2$ viewport in the dome, while the surface temperature is monitored via a FLIR infrared camera angled through a ZnSe window. The catalyst is purged in Ar for an hour before being reduced at 200 \textdegree{C }in a 5\% H$_2$ / 95\% Ar gas mixture for 2 hours. A simple Python script, the majority of which is seen in \cref{fig1}, is used to automate a series of temperature sweeps under varied laser powers and wavelengths, resulting in a 3-dimensional parameter sweep.  The initial data analysis, consisting of GC peak integration, species quantification, and conversion determination, is done fully using the built-in Catalight analysis tools described in the previous section and \cref{fig3}. The catalytic rates are calculated using a custom, specific script to produce Arrhenius plots such as those depicted in \cref{fig4}a for a fixed wavelength and variable power. The resulting apparent activation barriers deduced from a linear fit of the resulting rate data is plotted as a function of laser power and wavelength, as seen in \cref{fig4}b. 

Contrary to the results seen in plasmonic activation barrier measurements with different catalyst and chemistries, our results indicate an  \textit{apparent increase} in the activation barrier of the experiment in response to strongly and more resonant illumination of the system.\cite{Zhou2018} It has been suggested that plasmon-induced hot electrons could enhance the desorption of H$_2$ molecules from the surface of a catalyst.\cite{swearerAdvancingPlasmonInducedSelectivity2021, swearerHeterometallicAntennaReactor2016} As such, a plausible hypothesis could be that reduced H surface population leads to a slower, but more selective reaction rate on the surface of the catalyst. However, as seen in \cref{fig4}d and \cref{SI:C2H2_rate_order} and \cref{SI:H2_rate_order}, rate order measurement experiments on this catalyst for both H$_2$ and C$_2$H$_2$ show no change in response to light, with rate orders of approximately 1 and -0.5, respectively. The consistency of the rate order of each reactant in response to light does not support the hypothesis that the surface coverage of the catalyst has been altered by illumination. Additionally, it is plausible to hypothesize a change in the chemical selectivity for a partial hydrogenation reaction in response to a decreased H surface population, but our experiments did not show a noticeable increase in C$_2$H$_4$ selectivity under any tested reaction conditions (\cref{SI:X_vs_S}).

The number of studies available focused on the chemical selectivity of acetylene hydrogenation under plasmon excitation is limited. Previous research on this specific reaction utilized pure Pd nanoparticles supported on plasmon-active Al nanoparticles in an antenna/reactor configuration.\cite{swearerHeterometallicAntennaReactor2016} In this case, the active phase, support structure, and plasmon properties differ significantly from those of the system studied in our work. AuPd is known to undergo atomic scale restructuring, particularly at the surface, under different environmental conditions. The AuPd catalyst studied in this work was clearly deactivated at elevated reduction temperatures, a feature linked to Pd diffusion into the bulk of the catalyst.\cite{luneauEnhancingCatalyticPerformance} The use of lower temperatures supports the claim that Pd is still present on the surface of the active phase. However, support/catalyst interaction could certainly play a role in the differences observed. Furthermore, transient, atomic-scale diffusion of the alloy constituents could have a more complex effect on the behavior. The optical resonances of Au and Al nanoparticles differs, and plasmon resonances at different energy levels could cause different interactions between hot carriers and surface adsorbate between the two catalytic systems. The difference between results underpins a need to larger scale catalytic studies on a wide variety of systems, vindicating the development of this open source automation system.

As an alternative explanation to a plasmonic, hot-carrier driven reaction, we considered the possibility that the reaction is primarily driven by photothermal heating. In such a case, it is critical to consider the non-uniform temperature distribution of the catalyst pellet under illumination. Because the intensity profile of our laser is Gaussian-like, the resulting temperature profile is similarly non-uniform, and because the resulting rate is exponentially dependent on the temperature, it is inappropriate to assume a single-valued temperature when performing an Arrhenius fit. To explore the impact of this inaccuracy on our fitted data, we simulated the reaction rate of a theoretical 2D catalyst pellet using the experimentally determined dark activation barrier along with an approximated Gaussian 2D temperature profile. The 2D temperature profile was sectioned into infinitesimal polar sectors and the expected rate at each temperature sector was calculated using the dark activation barrier with no consideration given to a non-thermal reaction pathway. The average rate was then determined by averaging across spatial sectors of the catalyst, as opposed to assuming the average rate is similar to the rate induced by the average (or maximum) temperature of the catalyst pellet. The simulated reaction rate data is then fit against the maximum temperature of the catalyst pellet in a way which mimics the analysis performed on the experimental data. The resulting activation barrier map shows a clear artificial increase in the observed activation barrier, as seen in \cref{SI:Theory}. This increased barrier is an analysis artifact of fitting a reaction rate, which results from a complex temperature distribution, to a singular temperature value. Performing this same analysis on a much tighter temperature distribution than the experimental beam profile results in an even larger activation barrier increase, suggesting that the effect is most pronounced when there is a large degree of temperature inhomogeneity in the sample. Furthermore, the magnitude of the activation barrier increase for the experimental data is substantially larger than that of the simulation. A possible source for this discrepancy could arise from inaccuracy of the IR temperature measurement. An increased temperature in the simulation does show an increased activation barrier, but the maximum temperature of the sample would need to be 200 \textdegree{C} higher than the observed temperature to account for such an increase in the activation barrier. Additional inhomogeneity along the depth of the catalyst pellet could also be an additional source of error since the barrier increase is observed to be more pronounced for a tightly focused temperature profile. However, the depth of the catalyst bed (<1 mm) used here is substantially smaller than the width, so the radial temperature profile likely accounts for the dominant amount of inhomogeniety. 

Finally, this increased activation barrier could be explained by a partial photothermal effect combined with a photochemical/hot electron driven mechanism as previously hypothesized. While experimental results have shown decreasing activation barriers upon illumination, there have been theoretical predictions of increased activation barriers for hydrogenation chemistry on plasmonically excited Pd-based catalysts. Yuan et al showed that specific steps in the NH$_3$ dehydrogenation process on a Pd surface could be hindered by hot electron contributions.\cite{yuanPlasmonicPhotocatalysisChemically2022} This was further detailed by Wen et al. \cite{xuelanPlasmonDrivenAmmonia2024} Furthermore, Huang et al reported decreased hydrogenation rates for styrene on core@shell AuNR@Pd structures under conditions that favored the production of hot electrons.\cite{huangUnravelingSurfacePlasmon2016} In summary, photothermal simulations demonstrate that a slight increase in the observed activation barrier would be expected simply from the way in which temperature inhomogeneties are represented in an Arrhenius analysis. However, the activation barrier observed in our experiment is higher than the photothermal simulation. Combined with both theoretical prediction and experimental observations of hindered hydrogenation chemistry on different systems, it is plausible that this increased activation barrier is partially caused by hot-electron effects.

\section*{Conclusion} \label{conclusion}
The work presented within this report simultaneously introduces a new and powerful photocatalysis automation toolbox, demonstrates its effectiveness through the careful study of a meaningful plasmonic reaction, and presents unexpected but informative experimental results. A primary theme of this article is to introduce the concept of the Catalight code package to the community such that they may explore and consider using and developing it more for themselves. Much more information about the software tools, including a user guide, code documentation, and examples, is provided on the Catalight website.\cite{catalight_rtd} In addition to introducing the package to the community as it exists today, we also want to ask the community for help in developing this project further and provide some perspective on future possibilities. The ability to automate experiments is certainly desirable and could save researchers valuable time, but these tools can only be utilized with the development of more equipment modules. While many independent developers and hardware providers already produce APIs for their equipment, these will need to be “translated” to “speak” with Catalight. Guides for how to do this are written on the webpage. In addition, many users will want to develop and utilize different types of equipment. For example, GC analysis is currently supported and FTIR measurements are under development; however, future users may wish to add characterization techniques like mass spectrometry. New equipment types will require new abstract classes in addition to specific classes for each vendor. 

Entirely new experiment types could be integrated in the future as well. For example, the Catalight system could be easily expanded to function with photo(electro)chemistry with the addition of potentiostat drivers. Another exciting future direction could be the integration of data analysis with experiment execution in real-time. For example, rather than the user provide a predetermined time to wait for steady state, measurements could be made and analyzed during collection, and the software could be made to mathematically determine when steady state has been reached for itself. This type of implementation makes more sense for faster temporal resolution techniques such as online mass spectrometry compared to GC. If real-time analysis could be integrated with experiment performance, experiment design principles possibly combined with machine learning techniques could be introduced as well to guide optimization over wide parameter spaces. While many studies to date focus strongly on mechanistic understanding, future experiments could seek to optimize photocatalyst performance within this huge parameter space, making optimization techniques desirable.

Finally, the results of this work go beyond a simple demonstration of the reported automation package, and highlight some critical issues in specific plasmonic photocatalysis characterization techniques. We performed automated apparent activation barrier measurements on a Au$_{94}$Pd$_6$ plasmonic catalyst supported on $\gamma Al_2O_3$. We observed an increase in the activation barrier with increased laser power and photon energies. No change to the reaction rate order with respect to either reactant was observed. We noticed a spatial temperature distribution on the surface of our catalyst in response to light. Using simple thermal reaction calculations, we showed that a portion of this increased apparent activation barrier could be explained by incorrectly fitting an Arrhenius plot to a single surface temperature value when the actual surface temperature is non-uniform. Still, these results do not quantitatively match with our experimental observations, promoting further investigation of similar hydrogenation reactions on plasmonic particles. Whereas previous reports showed an improvement in the reaction behavior, namely through increased chemical selectivity, this report suggests that plasmonic excitation of this specific catalyst is dominated by photothermal effects and may even hinder the hydrogenation reaction. Between the limited number of plasmon-driven acetylene hydrogenation studies, a large number of parameters were changed. Further research is needed to understand which parameters (alloy composition, support material, plasmon frequency, etc) have meaningful impacts on the differences observed between different plasmon catalysis studies.

\clearpage

\section*{Acknowledgements}
The authors would like to acknowledge Mac Logan and Dylan Spagnuolo for helpful discussions involving Python best practices. 
This work was funded with support from Keck Foundation (Grant No. 994816), the National Science Foundation, Center for Adopting Flaws as Features (NSF CHE-2124983), and the Department of Energy, Office of Basic Energy Sciences, Division of Materials Science and Engineering (DE-AC02-76SF00515). B.B.B. was supported by the National Science Foundation Graduate Research Fellowship (DGE-1656518). Part of this work was performed at the Stanford Nano Shared Facilities (SNSF), supported by the National Science Foundation (ECCS-2026822). W.-H. Cheng acknowledges the generous support provided by the National Science and Technology Council, Taiwan (NSTC 113-2628-E-006-001) and the Ministry of Education, Taiwan (MOE-110-YSFEE-0005-002-P1).

\section*{Conflict of Interest}

The authors declare no conflict of interest.

\begin{shaded}
\noindent\textsf{\textbf{Keywords:} \keywords} 
\end{shaded}


\setlength{\bibsep}{0.0cm}
\bibliographystyle{Wiley-chemistry}
\bibliography{references}

\clearpage









\end{document}



\vspace{-1.5cm}\maketitle\vspace{-1cm}

\begin{figure} [!b]
\begin{minipage}[t]{\columnwidth}{\rule{\columnwidth}{1pt}\footnotesize{\textsf{\affiliation}}}\end{minipage}
\end{figure}




\section*{Sample Preparation}
\label{sample_prep}

\subsection*{Materials}
\label{materials}
Borane \textit{tert}-butylamine complex (TBAB), oleylamine (OLAM), Oleic acid (OLAC), and palladium (II) nitrate dihydrate (Pd(NO${_3}$)${_2}$${\cdot}$2H${_2}$O), were obtained from Sigma-Aldrich. 1,2,3,4-tetrahydronaphthalene (tetralin), hydrogen tetrachloroaurate (III) hydrate (HAuCl${_4}$${\cdot}$3H${_2}$O), reagent alcohol (90\% ethanol, 5\% isopropanol, and 5\% methanol), hexanes, and toluene were obtained from Fisher Scientific.  Alumina (Al${_2}$O${_3}$) was obtained from Sasol.

Each reagent in the following procedures is parenthetically followed by its stoichiometric amount, but for solutions to be transferred to a different container (e.g. the reducing solution in the Au NP synthesis), in practice a small excess of the solution was prepared for ease of transfer.

\subsection*{Synthesis of Nanoparticles (NPs)}
\label{synthesis}
\subsubsection*{Synthesis of Au NPs}

Au nanoparticles were synthesized using a modified version of a previously reported procedure\cite{cargnelloSynthesisColloidalPd, sunfacilesynthesismonodisperse}, targeting spheres with a diameter of 3-5 nm.  The synthesis was performed on 0.025, 0.23, and 0.5 mmol Au scales with similar results, though the smaller scales produced particles with a slightly narrower size distribution.  A representative procedure is described below for a 0.5 mmol (100 mg) Au scale synthesis:

First, the reducing agent TBAB (87 mg, 1.0 mmol) was dissolved in a mixture of OLAM (2 mL) and tetralin (2 mL), aided by sonication. A separate solution of OLAM (20 mL) and tetralin (20 mL) was prepared in the eventual reaction container, left at room temperature ($\approx$ 22 \textdegree{}C) to synthesize 5-nm diameter nanoparticles, or heated to 40 \textdegree{}C to synthesize 3-nm diameter nanoparticles.  The next two steps involving HAuCl${_4}$${\cdot}$3H${_2}$O were performed as quickly as safely possible to minimize exposure of the Au salt to air and moisture until reacted with TBAB.  HAuCl${_4}$${\cdot}$3H${_2}$O (200 mg, 0.5 mmol) was dissolved in the reaction container, also aided by sonication and magnetic stirring, to form a dark orange solution. As soon as the Au salt was fully dissolved, the entire TBAB reducing solution was quickly injected into the orange Au salt solution while stirring to initiate a burst nucleation, immediately turning the solution dark purple. This resulting reaction solution was stirred in air at the same reaction temperature for 1 hr.

The Au NP were recovered via multiple precipitation/washing cycles.  Although the optimal selection and amounts of wash solvents have been observed to vary from lab to lab, we report our most consistent conditions here. For the first wash, the reaction solution (44 mL) was apportioned into aliquots in separate centrifuge tubes (one 50 mL tube for each $\approx$ 11 mL of solution). To each tube, reagent alcohol (20-30 mL) was added as the wash solvent to precipitate out the Au NP while keeping any unreacted reagents dissolved, followed by centrifugation (5000 g, 3 min) to recover the NP.  The clear supernatant was discarded, then the Au NP pellet in each tube was redispersed in a minimal amount of hexanes ($\approx$< 1 mL each).  Toluene can be used as an alternative to hexanes to minimize evaporation.  For the 2nd precipitation/washing cycle, reagent alcohol (20-30 mL) was again added to each tube followed by centrifugation (5000 g, 3 min).  The Au NP are observed to be less prone to precipitation after the 2nd wash, so a 3rd cycle is optional if greater particle cleanliness is desired.  To fully precipitate the entire yield of Au NP, more reagent alcohol (40 mL) was used as the wash solvent, followed by centrifugation at 10000 g for 30 min. We note that the centrifugation speeds necessary to precipitate out all of the NP in the 3rd cycle may cause particle agglomeration, and so the speed can be decreased to preserve NP morphology at the cost of final yield. The final particles were redispersed in a total of 5 mL of toluene for long-term storage.


\subsubsection*{Synthesis of AuPd NPs}

AuPd nanoparticles were synthesized using a modified version of a previously reported seed-mediated procedure\cite{cargnelloSynthesisColloidalPd, murrayUniformBimetallicNanocrystals}, targeting spheres with a diameter of 5 nm and, a homogeneous surface alloy of AuPd, and a total nominal Pd:Au ratio of 5:95.  We note this atomic ratio is given for the nanoparticle as a whole, despite the Pd being expected to only alloy in the surface.

A solution of Au NP (30 mg) in hexanes or toluene (synthesized and washed as described above) as seeds, OLAM (30 mL), OLAC (1.9 mL), and Pd(NO${_3}$)${_2}$${\cdot}$2H${_2}$O (3.8 mg) were mixed in a three-neck flask (100 mL).  This reaction solution is kept under constant magnetic stirring.  The solution was setup on a Schlenk line and degassed at 50 \textdegree{}C under vacuum for 30 minutes (also removing hexanes/toluene and any residual water).  The solution was then placed under flowing inert gas (nitrogen), heated to 140 \textdegree{}C, and held at that temperature for 30 min, after which it was cooled to room temperature.

The AuPd NP were recovered by precipitation/washing in a nearly identical procedure to the Au NP described above, with the difference that a small amount of OLAM ($\approx$10 \textmu{}L) was sometimes added with a particular hexanes/toluene addition to aid in resuspending the particles.  A similar amount of OLAM can also be added to the final solution to prevent the AuPd NP from precipitating out during long-term storage.

The final size of the AuPd NP is determined by the size of the Au NP seeds, while the relative Pd composition can be increased by adding more Pd(NO${_3}$)${_2}$${\cdot}$2H${_2}$O to the reaction.  Increasing the amount of Pd can further increase the final size of the AuPd NP based on the additional material added.  We observe that typically, the final Pd:Au ratio in the NP is lower than the molar ratio of the added reagents (Pd salt and Au seeds, respectively).

\subsection*{Loading of Nanoparticles onto Support}

Nanoparticles were loaded onto an Al${_2}$O${_3}$ support using a modified version of a previously reported procedure\cite{cargnelloSynthesisColloidalPd}, aiming to disperse individual NP throughout the surface of Al${_2}$O${_3}$ microparticles.  The Al${_2}$O${_3}$ powder was previously calcined in air at 900 \textdegree{}C overnight.  Here, we describe a typical procedure targeting a nominal 4 wt\% loading of NP.

Al${_2}$O${_3}$ powder (200 mg) was first dispersed in hexanes/toluene (2 mL) via sonication for 5 min.  This mixture was then vigorously stirred while a NP (8.3 mg of metal, as determined by thermogravimetric analysis) solution was added drop-wise.  The NP-support mixture was stirred for 15 min, then centrifuged (5000g, 3 min).  We note that transferring of the wet powder to a centrifuge tube can incur significant yield losses, which should be kept in mind in deciding the original scale.  If the supernatant after centrifugation was not clear, which is more likely to occur with NP solutions with excess OLAM (either remaining from the original reaction solution or added back in after the wash), then reagent alcohol (20 mL) was added as an anti-solvent and the mixture was centrifuged again (5000 g, 5 min).  The supernatant was then discarded, and the remaining solids dried overnight at room temperature if the solvent was hexanes, or in a drying oven at 60 \textdegree{}C if the solvent was toluene.

This dried powder was then ground and sieved through a 170 mesh (90 \textmu{}m) sieve.  The sieved powder was collected in an alumina crucible and placed in a furnace, where it was calcined in air at 300 \textdegree{}C for 30 minutes to remove all the organic ligands surrounding the NPs and activate the catalyst.

The support-loading steps are largely agnostic of the alloy composition of NP being loaded, but we noticed that our NP solutions containing more Pd are more likely to have required the addition of small volumes of OLAM to keep them suspended, and therefore they would more likely require addition of the extra anti-solvent during the centrifugation step.

\subsection*{Reactor Set-up and Catalyst Pretreatment}
\label{pretreatment}

\begin{figure}
\begin{center}
    \includegraphics[]{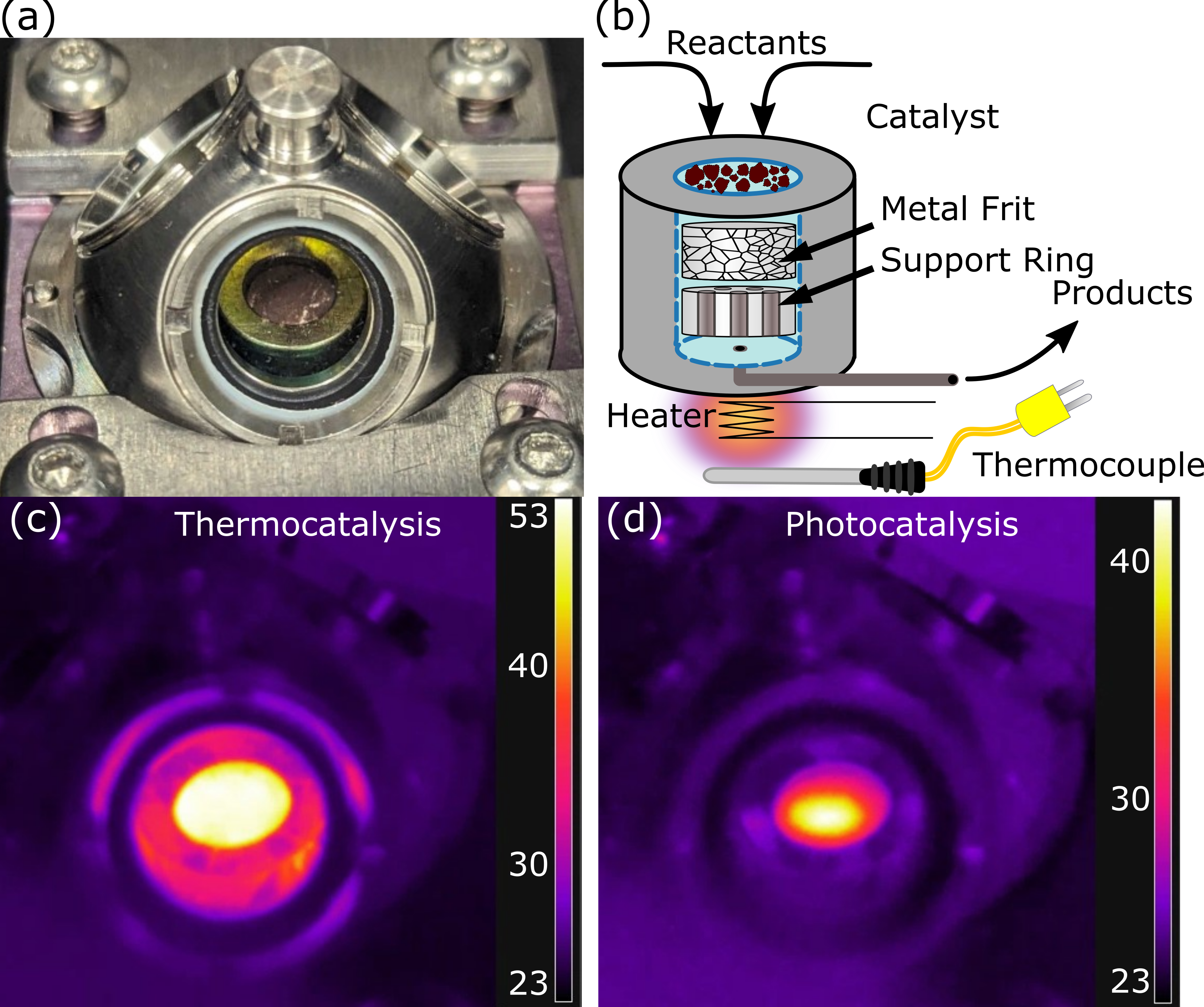}
    \caption{Overview of reactor configuration used in photocatalysis measurements. (a) photograph of the loaded reactor viewed from the illumination window. (b) schematic of the reactor cup showing the supporting metal frit, support ring, gas flow pathway, thermocouple, and resistive heater. (c) IR camera image of the reactor temperature distribution during thermal catalysis. (d) IR camera image during photocatalysis.}
  \label{SI:schematic}
\end{center}
\end{figure}

The catalyst sample is placed into a Harrick high temperature reaction chamber outfitted with a 3-window, FTIR compatible dome, as depicted in \cref{SI:schematic}a. The active sample (metal-loaded) depth is minimized given the penetration depth of light is low through the Al$_{2}$O$_{3}$ support material. A custom made stainless steel puck is placed in the bottom of the sample compartment to improve the mechanical stability of the mesh sieve. The puck is 1 mm thick and 6.15 mm in diameter with 7 hexagonally arranged, 1.3 mm holes cut-out from the puck to allow airflow. The center-to-center distance between holes is 1.8625 mm. A 325x325 mesh made of 316 stainless steel is placed atop the support puck to stop the flow of particulates. In the activation barrier experiments, pure Al$_{2}$O$_{3}$ support material is placed in the reactor to raise the level of the active component such that is is flush with the top of the reactor chamber. In later experiments, a 1/4" diameter x 1/16" thick porous stainless steel disc designed to filter 40 \textmu m particles was added in place of the pure Al$_{2}$O$_{3}$ and proved to be easier to handle. The disc was sanded slightly to fit tightly within the reactor chamber. Finally, 2 to 4 mg of metal-loaded catalyst is placed on the top surface, enough to form a cohesive layer, and the reactor system is sealed. The sample chamber layout is depicted in \cref{SI:schematic}b. The sample temperature is controlled using a built-in resistive heater on a feedback loop with a thermocouple located a few millimeters beneath the sample. The surface temperature of the catalyst is monitored using an IR thermal camera as depicted in \cref{SI:schematic}c \& d. In a thermal reaction (\cref{SI:schematic}c) the sample is evenly heated and a substantial temperature increase in the outer walls of the sample cup is observed. In a photocatalytic reaction (\cref{SI:schematic}d), the sample heating is spatially non-uniform with the highest temperatures observed near the center of the sample.

\begin{figure}
\begin{center}
    \includegraphics[]{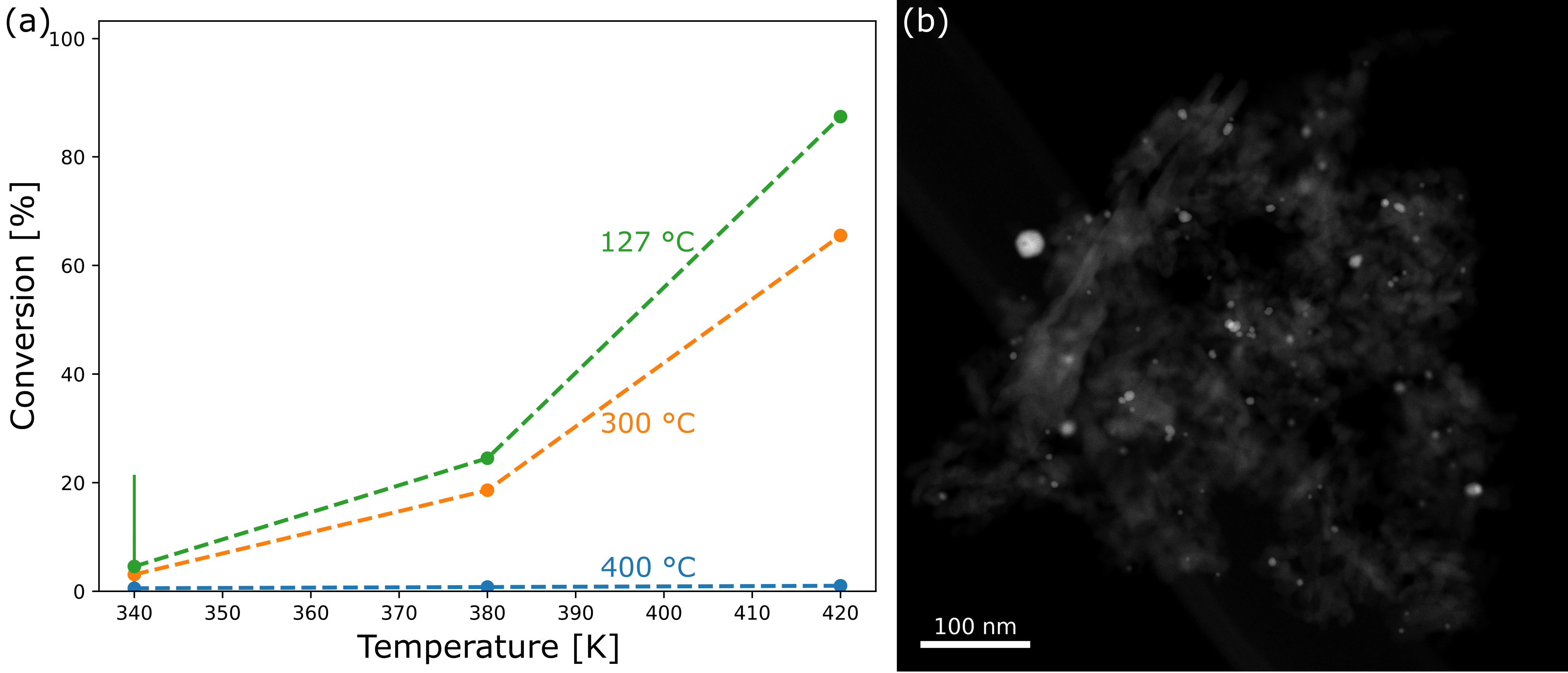}
    \caption{The effect of reduction temperature on the performance of a Au$_{94}$Pd$_6$ catalyst. (a) The sample is treated at the indicated temperature (127, 300, or 400 °C) for 2 hours then C$_2$H$_2$ conversion is measured through a temperature sweep from 340 K to 420 K. (b) STEM image of Au$_{94}$Pd$_6$ sample after reduction at 400 °C.}
  \label{SI:reduction}
\end{center}
\end{figure}

The catalyst is flushed with Ar for >1 hr to remove residual gasses prior to the reduction and reaction steps. The sample is reduced in a 5\%/95\% H$_2$/Ar environment to remove any oxide formed during the calcination procedure. The reduction temperature is observed to have a significant impact on the material's activity, as seen in \cref{SI:reduction}a. When reduced at 400 °C for two hours, the sample is rendered almost fully deactivated. Treatment at 127 °C or 300 °C both show reasonable activity. \cref{SI:reduction}b shows a STEM image of the sample after reduction at 400 °C. Some sintering is observed in the sample, but there are also reports of Pd becoming unstable on the catalysts surface under H$_2$ treatment at elevated temperatures.\cite{luneauEnhancingCatalyticPerformance} These findings are directly in line with those reports.

\begin{figure}
\begin{center}
    \includegraphics[]{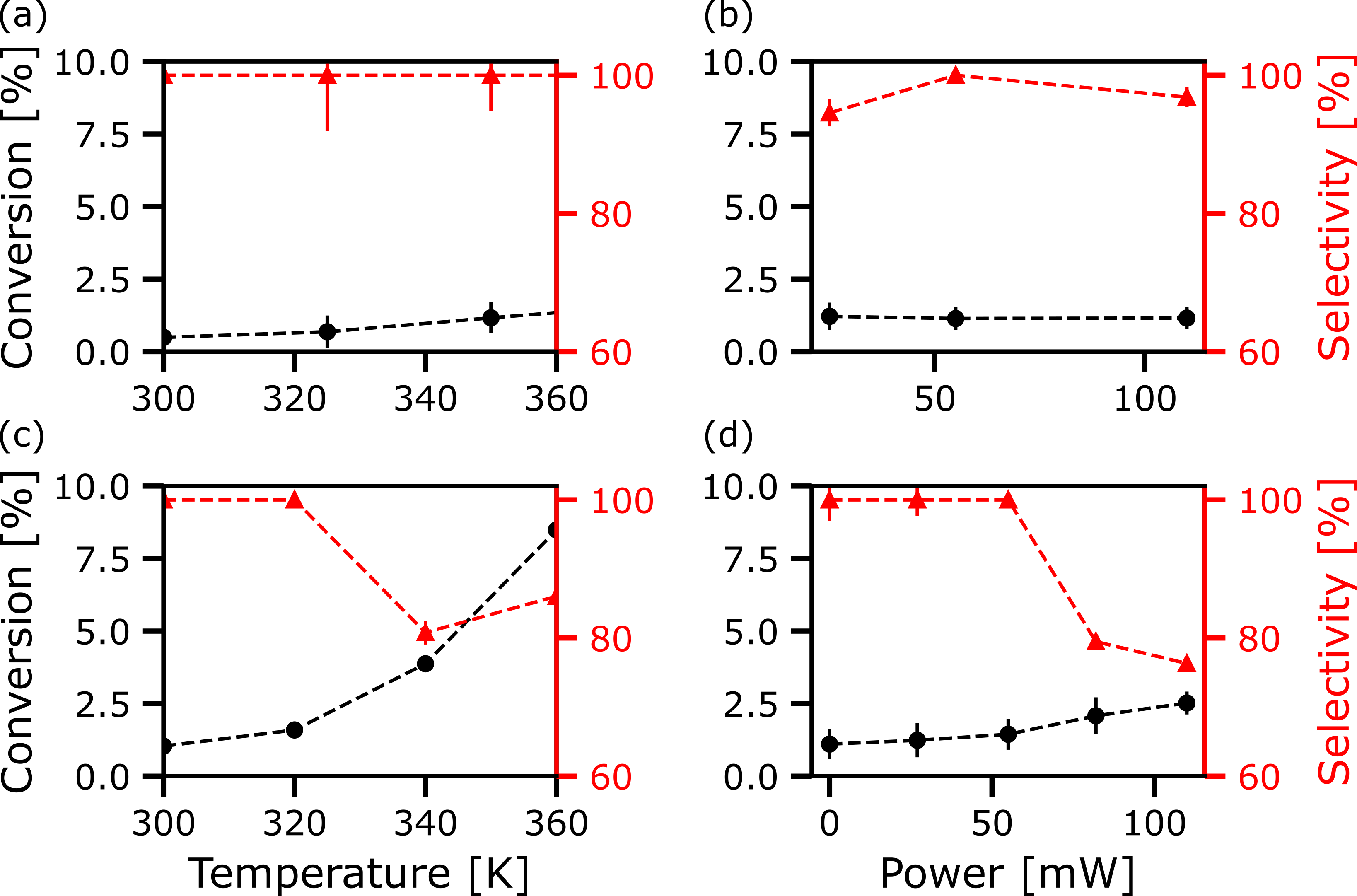}
    \caption{Thermal and photo reaction on Au and Au$_{94}$Pd$_6$. (a) Thermal and (b) photo reaction on pure Au. (c) Thermal and (d) photo reaction on Au$_{94}$Pd$_6$. All reactions were carried out on 2.5 mg of catalyst at 10 sccm using 1\% C$_2$H$_2$ and 5\% H$_2$}
  \label{SI:Au_vs_AuPd}
\end{center}
\end{figure}

\cref{SI:Au_vs_AuPd} shows a simple comparison between thermal and light-driven reactions on pure Au and Au$_{94}$Pd$_6$ catalysts. The pure Au catalyst shows very little reactivity in comparison to the alloy. While the selectivity of the catalyst appears higher, the conversion is much lower. For the same conversion shown in the Au$_{94}$Pd$_6$ catalyst, the selectivity is also very near 100\%. The light driven reaction on the pure Au catalyst shows no response in conversion, though the selectivity of the reaction does vary slightly. In comparison, the Au$_{94}$Pd$_6$ catalyst shows a response thermally and optically. The activity for the thermal reaction is higher than the light driven reaction under this given set of conditions, but it is not meaningful to compare the maximum conversion for either reaction because we have no good means of relating the overall temperature of the catalyst to the optical input. 

\begin{figure}
\begin{center}
    \includegraphics[]{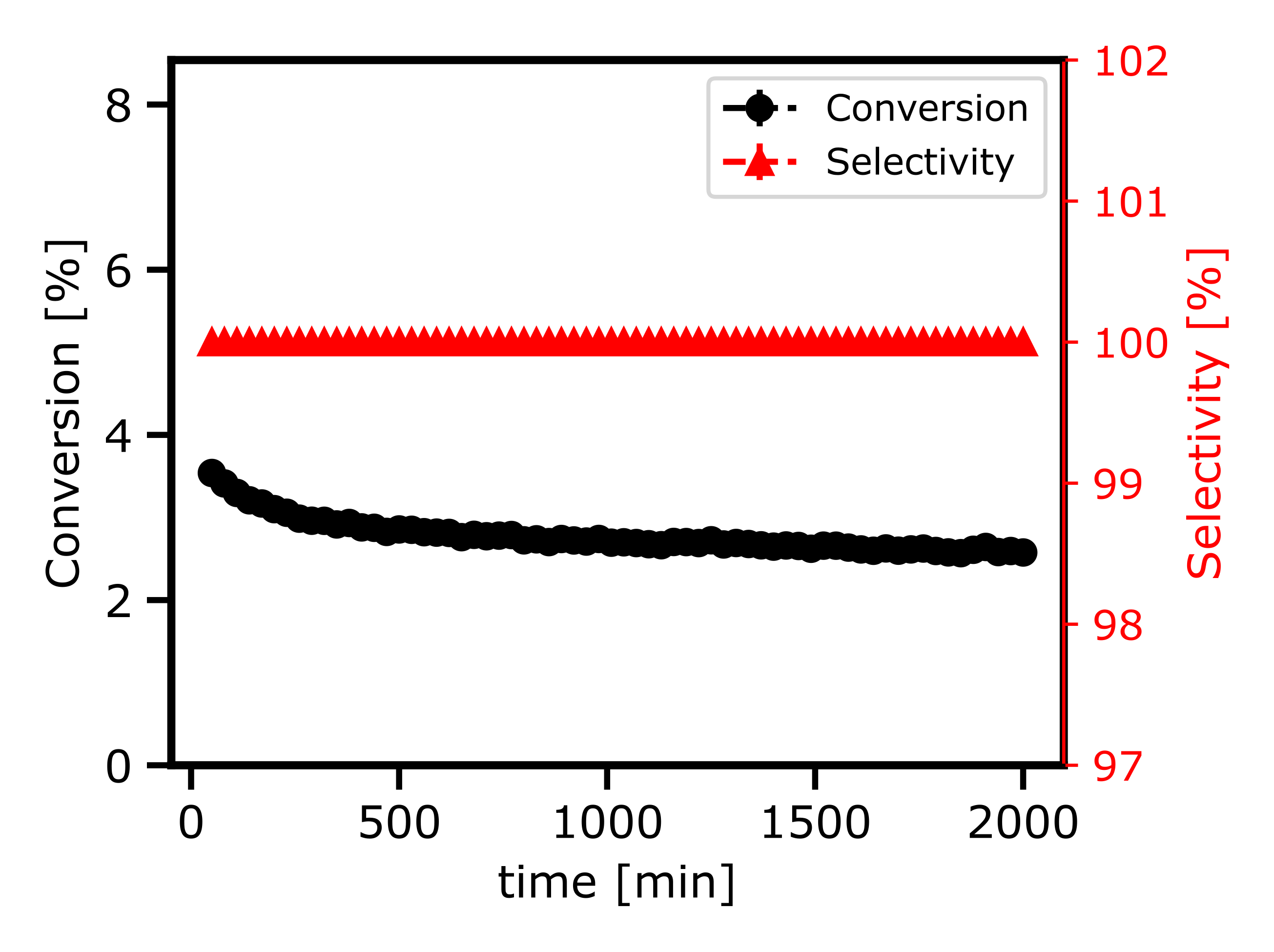}
    \caption{Stability test of Au$_{94}$Pd$_6$ catalyst at 380 K and 50 sccm, mimicking the most harsh reaction conditions of the extended activation barrier measurement experiment}
  \label{SI:stability}
\end{center}
\end{figure}

\cref{SI:stability} shows the stability of the Au$_{94}$Pd$_6$ alloy during the thermal reaction. Over the first few hours of operation, the reactivity of the catalyst drops steeply. After around 500 minutes, the catalyst reaches a more stable period.

\section*{Sample Characterization}

\subsection*{Compositional Analysis}

\begin{table}
    \centering
    \begin{tabular}{|c|c|c|}
    \hline
        Catalyst Sample & Pd:Au ratio & wt\% metal on Al$_2$O$_{3}$\\
        \hline
        Au & 0:100 & 3.0\%\\
        \hline
        Pd & 100:0 & 2.0\%\\
        \hline
        AuPd & 94:6 & 2.3\%\\
        \hline
    \end{tabular}
    \caption{Inductively coupled plasma mass spectrometry (ICP-MS) was performed to determine the ratios of Au, Pd, and Al$_2$O$_{3}$ in our catalyst sample, quantifying the Pd:Au ratio in the alloyed NPs and the mass of metal NP loaded on the Al$_2$O$_{3}$ support, expressed as a wt\% of the total catalyst that is Au/Pd.}
    \label{tab:S1}
\end{table}

\subsection*{Transmission Electron Microscopy}
\label{tem}

\begin{figure}
\begin{center}
    \includegraphics[]{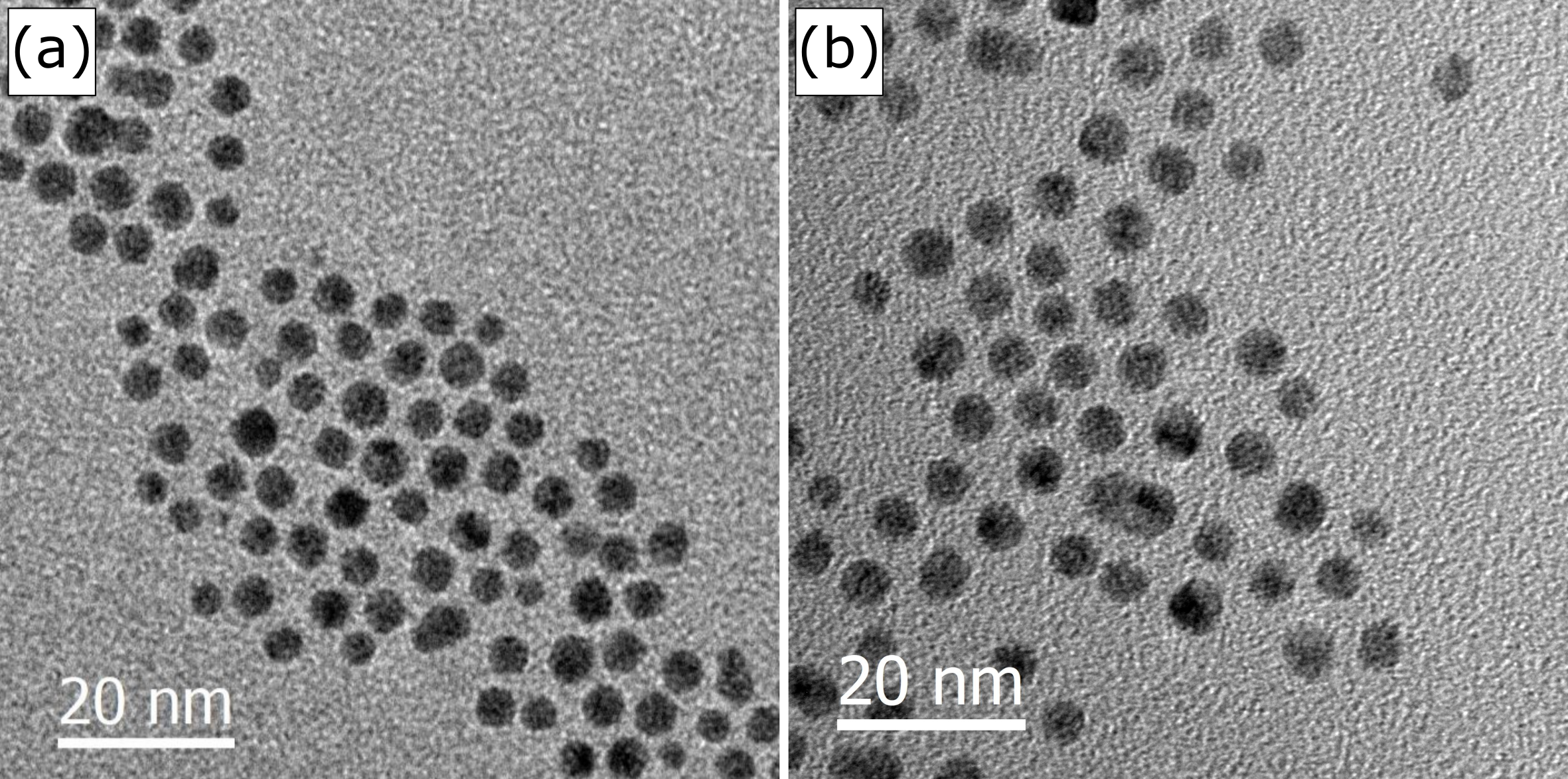}
    \\
    \includegraphics[height = 5cm]{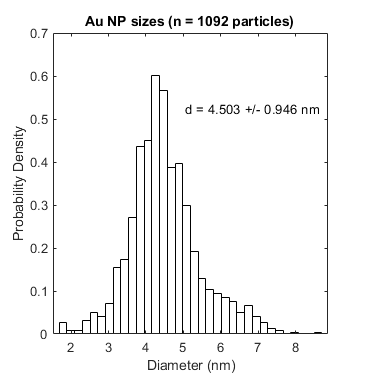}
    \includegraphics[height = 5cm]{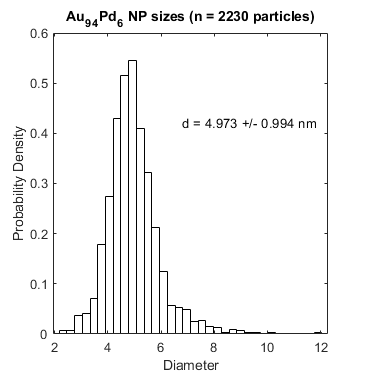}
    \caption{Representative transmission electron micrographs of $\approx$5-nm diameter Au NP (a) and Au${_{94}}$Pd${_6}$ NP (b) and their measured particle size distributions}
  \label{SI:TEM_colloidal}
\end{center}
\end{figure}

\begin{figure}
\begin{center}
    \includegraphics[]{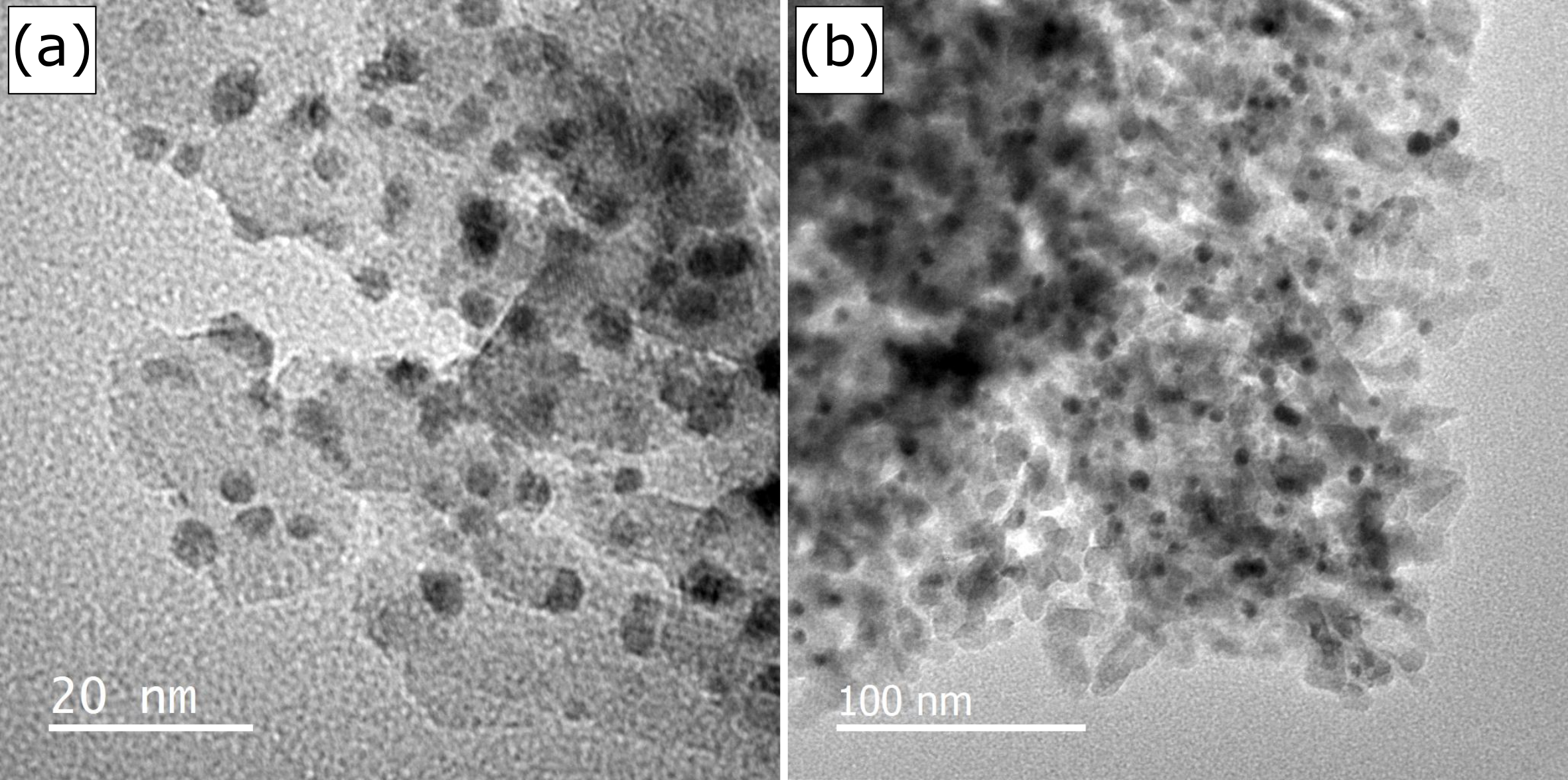}
    \caption{Representative transmission electron micrographs of Au NP (a) and Au${_{94}}$Pd${_6}$ NP (b) supported on Al$_2$O$_{3}$, after calcination}
  \label{SI:TEM_support}
\end{center}
\end{figure}

\begin{figure}
\begin{center}
    \includegraphics[]{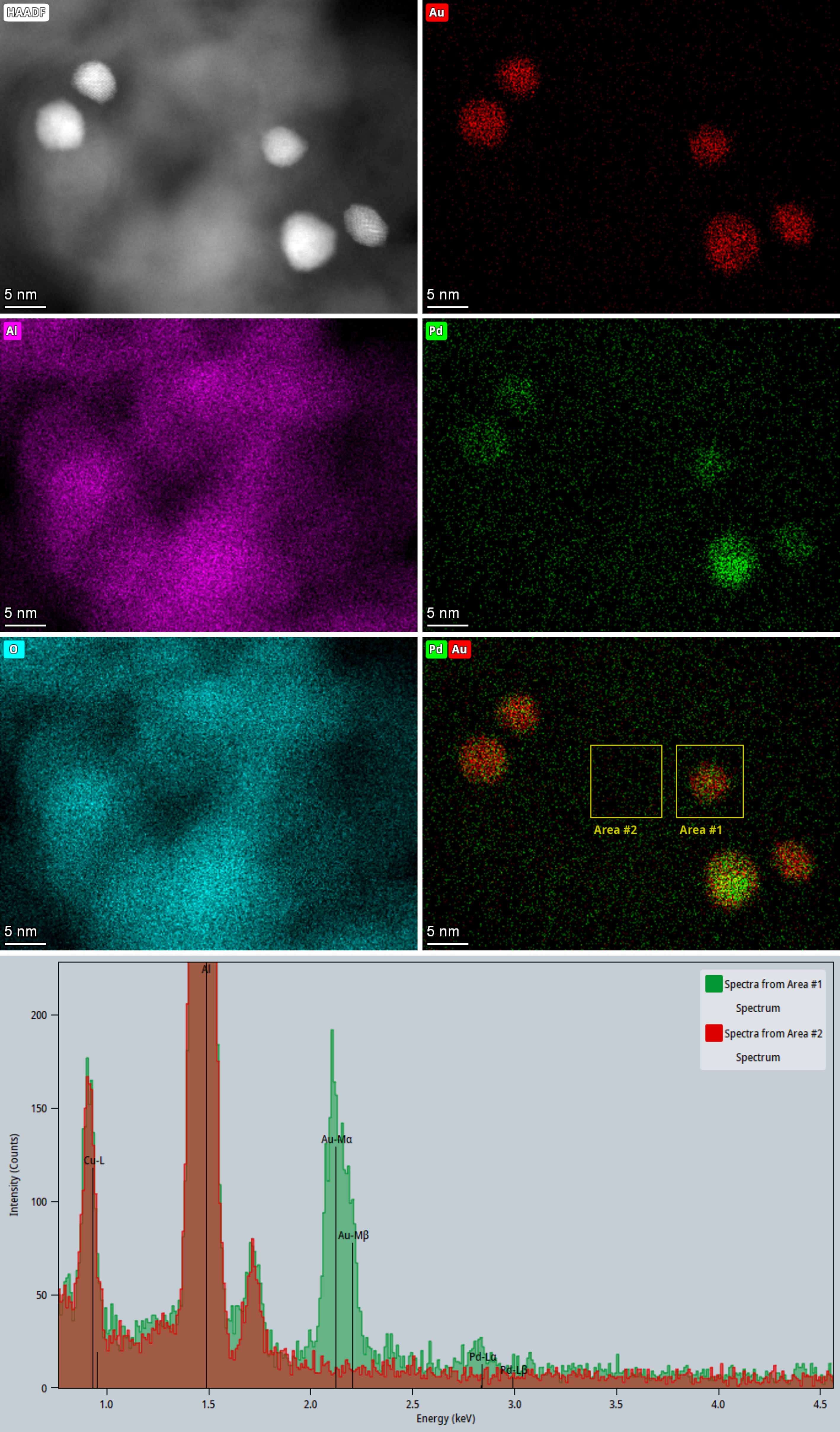}
    \caption{Scanning transmission electron micrograph and EDX spectral maps of the calcinated Au${_{94}}$Pd${_6}$/Al$_2$O$_{3}$ catalyst.  The high-angle annular dark field (HAADF) image (top-left) shows high Z-contrast AuPd nanoparticles on top of a fainter Al$_2$O$_{3}$ background.  The subsequent false-colored EDX maps shows the spatial distribution of each element. The last overlaid Au and Pd EDX map has selected areas corresponding to the spectra at the bottom, showing the presence of Au (M) and Pd (L) lines when on a nanoparticle (Area \#1) but complete absence of those peaks when on an adjacent area (Area \#2).}
  \label{SI:STEM_EDS}
\end{center}
\end{figure}

\subsection*{Optical Characterization}
\label{optics}
The optical extinction of the colloidal nanoparticle solutions was measured using a Cary 6000i UV/Vis spectrometer outfitted with an integrating sphere. The samples were placed in a quartz cuvette and diluted in toluene such that a the sample was transparent and only slightly colored to the eye. The sample was placed at the entrance of the integrating sphere and the transmitted light was recorded.

\begin{figure}
\begin{center}
    \includegraphics[]{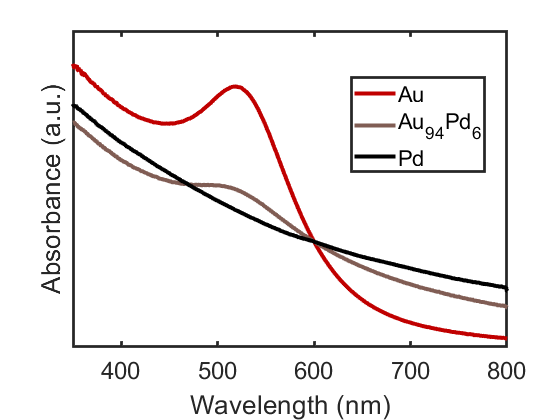}
    \caption{UV-Vis absorbance spectra of Au and AuPd NP in hexanes, showing a surface plasmon resonance peak around 520 nm which is dampened upon alloying of Au with Pd.}
  \label{SI:UVVis}
\end{center}
\end{figure}

\section*{Equipment and measurement setup}
\label{equipment}
\subsection*{Reactant Delivery}
\label{gas}
Controlled reactant gas flow is provided by a combination of 4 MC-Series Alicat mass flow controllers (MFCs). Gas supply lines are connected to the parallel MFCs using a series of 3-way selection valves such that the input gas can be down-selected from 9 distinct input gasses. N$_2$ and Ar are supplied to each MFC line such that the supply lines can be purged of other reactants prior to an experiment. 1/8" stainless steel tubing, valves and compression fittings are used throughout the system, except for the immediate connection to the reaction chamber which utilizes 1/8" PEEK tubing for a flexible connection. The 4 MFC lines converge into a single reactant supply line using a stainless steel gas manifold. The reactant supply line passes through a double-ended 10 mL sampling cylinder filled with stainless steel ball bearings to produce turbulent flow and enhance gas mixing. Stainless steel wool is used to fill the tube connection of the sampling cylinder such that the ball bearings cannot completely block gas flow on the outlet side. The reactant supply is connected to the reaction chamber via and optional by-pass loop and routed to the gas chromatograph (GC). A particle filter followed by a one-directional check valve are placed before the inlet of the GC to prevent accidental particle flow into the check valve and reverse flow during GC sampling, respectively. The output of the GC is monitored via an Alicat mass flow meter and the exhaust is routed to a chemical ventilation line.
\subsection*{Product Analysis}
\label{gc}
Product analysis is performed using an SRI 8610C gas chromatograph. The GC is outfitted with a flame ionization detector (FID) and a thermal conductivity detector (TCD) connected on independent auto-samplers. The FID sampler uses a custom 150 \textmu L sample loop while the TCD uses a 1 mL sample loop. Light elements are optionally detected using a 0.5 m Hayesep D pre-column followed by a 1.5 m Moleseive 5A column (not used in this work). The majority of elements are separated through a 2 m Hayesep N column to achieve separation of the C2 molecules. The FID is used on the "medium" gain setting while the TCD is not used in this work. The GC oven temperature is held at 40 \textdegree{C} for 1.5 min, ramps to 75 \textdegree{C} at 50 \textdegree{C}/min, held at 75 \textdegree{C} for 5 min, ramped to 145 \textdegree{C} at 60 \textdegree{C}/min then held at 145 \textdegree{C} for 12.63 min for a 20 min collection time.

\subsection*{Illumination}
\label{light}
Light excitation is achieved by sending collimated laser light through the "sample viewing port" of the reactor dome. The reactor is mounted inside of a Thermofisher Nicolet IS-50 FTIR (not used in this work), limiting physical access to the reactor from directly above. As such, laser light is sent at a roughly 40\textdegree{} angle from the front portion of the reactor. A custom-built laser collimation stage is mounted on a moving platform in-front of the FTIR such that the optics system can be moved entirely out of the way when replacing the reactor between sample loading. An achromatic doublet lens with 400-700 nm anti-reflection coating and a 7.5 mm focal length is used to collimate laser light directly out an SMA ended multimode fiber. The exact fiber is changed depending on the excitation laser. In this work, an FD10 fiber from NKT Photonics is used to couple in filtered laser light from a NKT FIU15 Fianium laser. The laser light is filtered using an NKT Varia variable band pass filter. Two mirrors are used to guide the light vertically, then down to a roughly 40\textdegree{} angle into the reactor. The collimated output is slightly defocused such that the beam diameter is roughly the size of the catalyst bed ($\approx{}$6.3 mm). The entire fiber collimating apparatus is mounted on an x-y translation stage, which is used to finely position the center of the beam into the center of the catalyst bed.

\subsection*{Surface Temperature Measurements}
\label{temperature}
\begin{figure}
\begin{center}
    \includegraphics[]{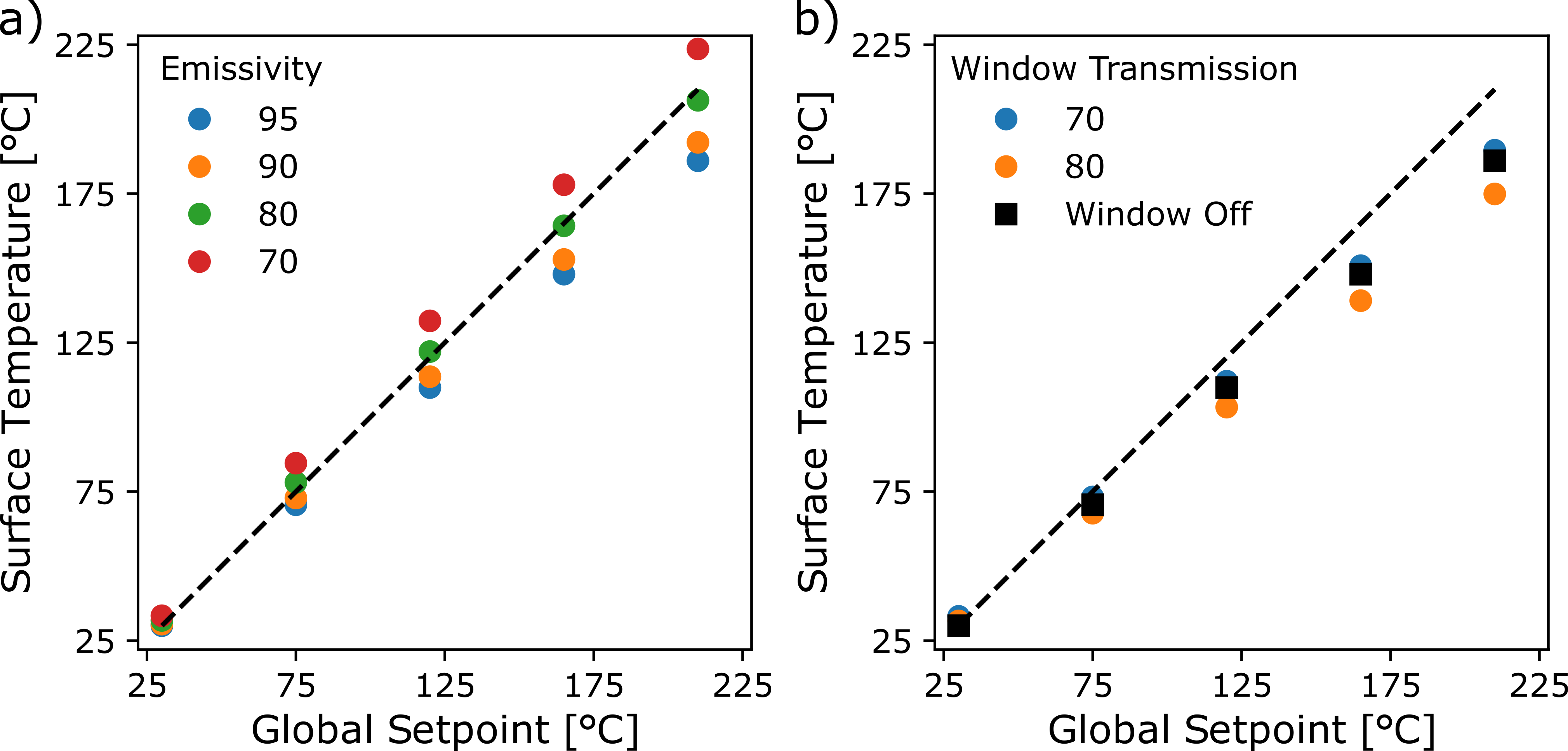}
    \caption{Surface temperature measurements of heated activated carbon under various camera settings. a) The surface temperature measured with the window removed using various emissivity settings. b) A change in the window transmission parameter as compared to windowless measurements using an emissivity of 0.95.}
  \label{SI:IR_Cam}
\end{center}
\end{figure}

Bulk sample temperature is controlled using the factory supplied Harrick heating system and calibration. This is used as both the resistive heat supply and to sense and set the global sample temperature. The sample surface temperature is monitored using a FLIR A700 thermal camera. The thermal camera is used with the factory calibration. The reactor sample compartment was filled activated carbon (emissivity $\approx{}$0.95) and the temperature was measured at various temperatures with the IR transparent dome removed to confirm the accuracy of the measurement. As seen in \cref{SI:IR_Cam}a, this showed a noticeable deviation from the global set point of 10-12\%. In \cref{SI:IR_Cam}b, the temperature was then measured with the IR transparent dome in place, and the transmission value in the FLIR software was adjusted to produce a consistent temperature reading to that which was recorded with the dome removed. This led to a empirically observed transmission value of 72\% which is slightly higher than the manufacturer's reported value. The camera is placed approximately 0.23 m from the reactor chamber and the inner diameter of the sample cup (6.3 mm) is used as a reference point for the spatial calibration of the camera. The emissivity of the catalyst sample is assumed to be 0.95 and produces consistent measurements to the activated carbon at this value. It should be noted that this assumed emissivity value has received significant skepticism in the literature, and a proper emissivity measurement would greatly improve the reliability of the results.\cite{sivanCommentQuantifyingHot2019} However, our measurements using this value are consistent with those taken for activated carbon, and an assumed emissivity of 0.95 is reasonable for activated carbon as it behaves as a black body material.

\subsection*{Activation Barrier Measurements}
\begin{figure}
\begin{center}
    \includegraphics[]{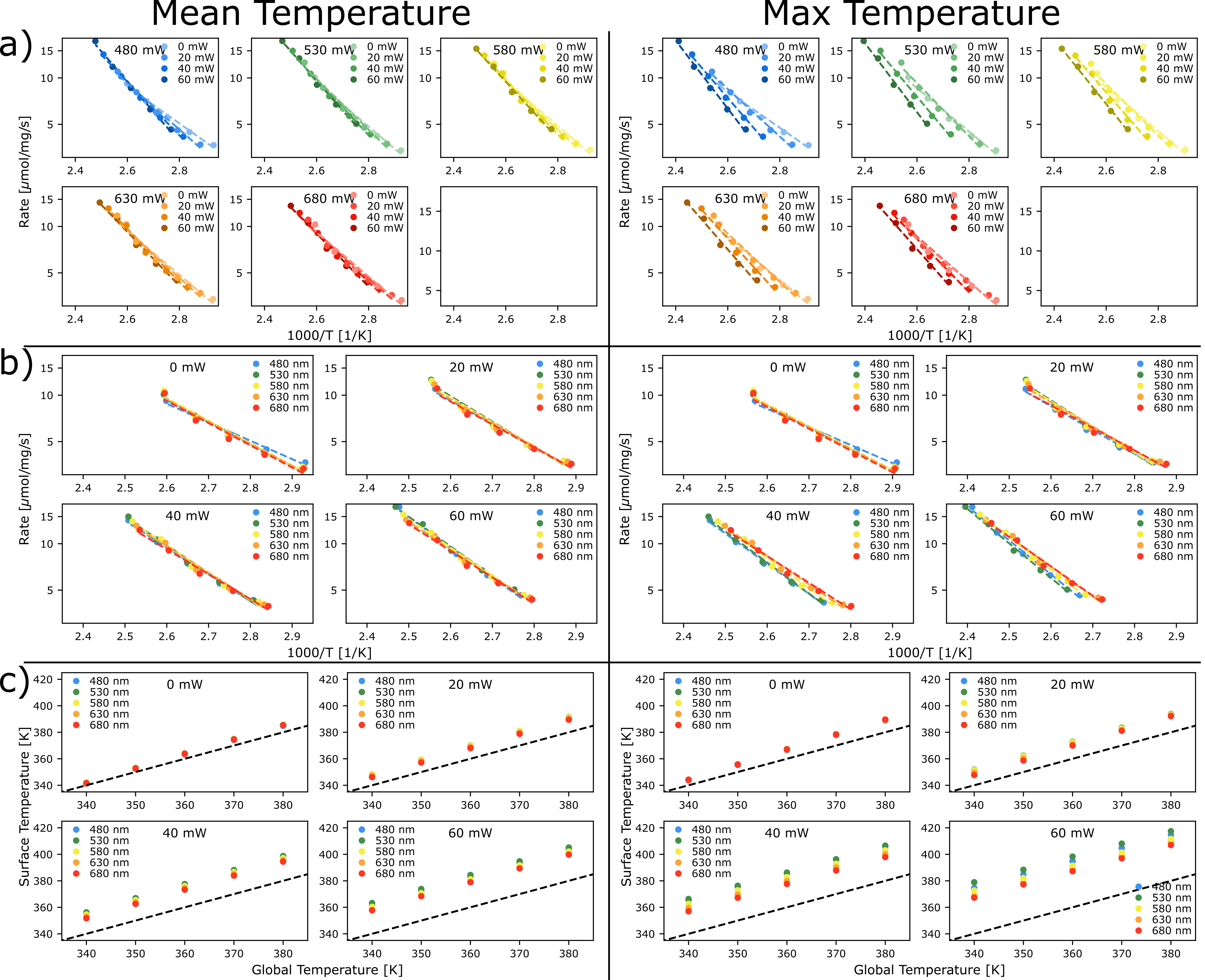}
    \caption{Comparison of the activation barrier experiment analysis using both the max and mean surface temperature. a) Arrhenius fit for various powers at a fixed center wavelength. b) Arrenhius fits for various wavelengths at a fixed power. c) the measured mean or max surface temperature as a function of the thermocouple set point.}
  \label{SI:ActivationBarrier}
\end{center}
\end{figure}

As described in the main portion of the text, reducing the inhomogenous temperature distribution of the catalyst to a single value results in fitting errors for the Arrhenius fitting procedure. \cref{SI:ActivationBarrier} displays the full fitting procedure for every data set using both the mean and max surface temperature of the catalyst. \cref{SI:ActivationBarrier}a shows the fitting procedure for various powers at a fixed center wavelength. The various in the the max temperature shows a greater distribution of fits for this dataset. \cref{SI:ActivationBarrier}b shows the same data, but displayed for various center wavelengths at a fixed power. The variation in the data fits is much less for all circumstances here. However, the surface temperature varies much more in each plot, which is most easily seen by comparing the x location of the plotted data for the 0 mW and 60 mW cases. The higher power case is clearly much higher in temperature. \cref{SI:ActivationBarrier}c shows the measured surface temperatures as a function of the global setpoint. The difference in the setpoint an observed surface temperature is relatively uniform as a function of the global temperature, but the difference increases with higher power and shorter wavelength.

\subsection{Theory}
\begin{figure}
\begin{center}
    \includegraphics[]{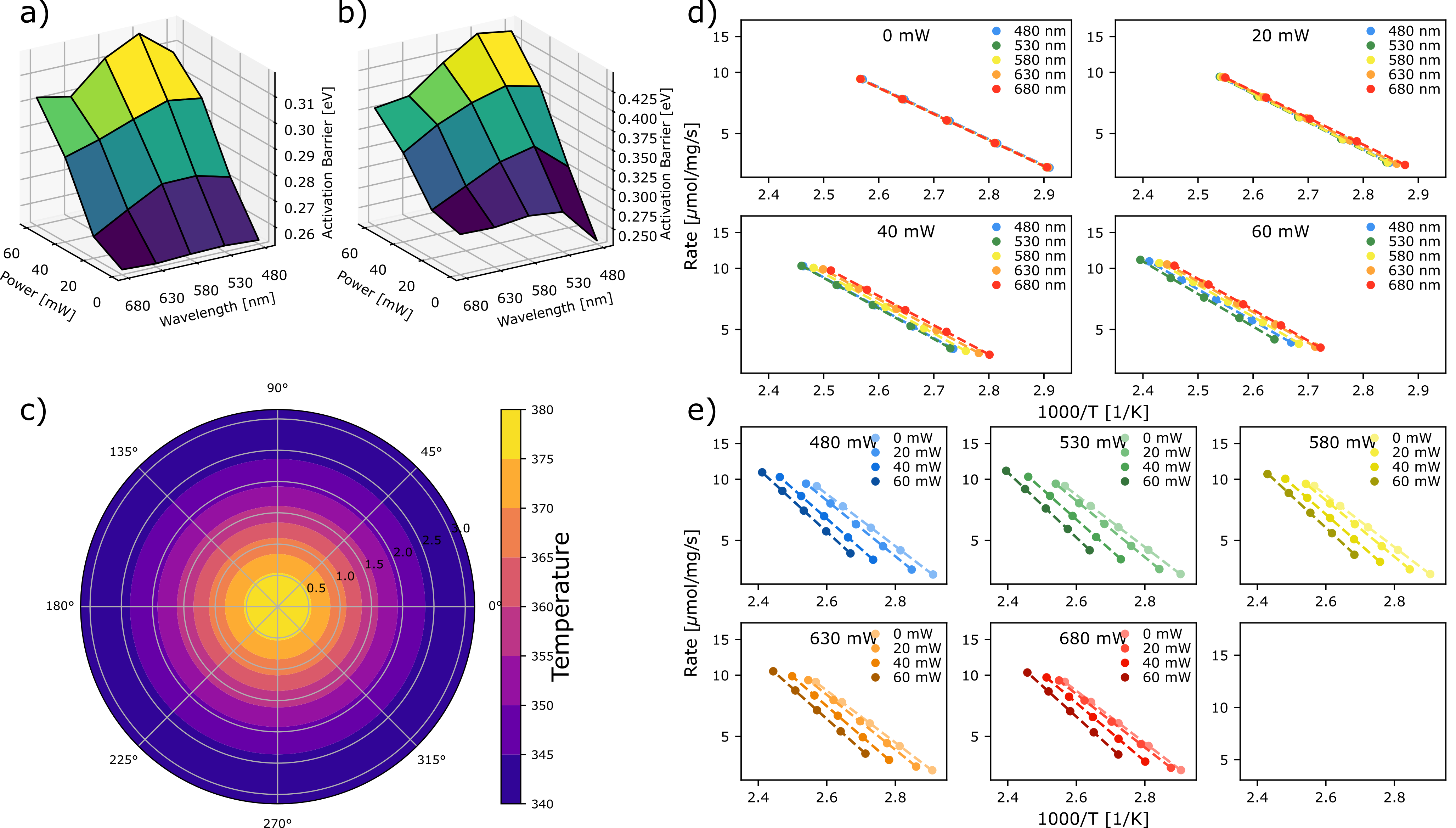}
    \caption{a) Simulated activation barrier assumming a purely photothermal mechanism. b) The experimentally observed apparent activation barrier. c) The simulated photothermal distribution for a baseline temperature of 340 K and 530 nm illumination at 60 mW. Arrhenius fitting plots using the maximum temperature of the surface for d) fixed powers and e) fixed center wavelengths}
  \label{SI:Theory}
\end{center}
\end{figure}

The effects of an inhomogeneous temperature distribution on the Arrhenius fitting procedure was tested by simplistic simulations assuming a purely thermal reaction mechanism. The activation barrier for the thermal reaction is determined using the average activation barrier measured in the light off scenario. An artificial 2D temperature distribution is created by superimposing the global temperature setpoint (340-380 K) with a Gaussian temperature distribution with a 5 mm full width tenth max (approximating the laser beam dimensions) The maximum temperature of this Gaussian is set to be the maximum temperature observed within the real experiment. An example of this simulated temperature distribution is shown in \cref{SI:Theory}c. This simulated temperature distribution is used to determine the effective thermal rate of the reaction at 1e6 different points in space assuming a uniform temperature along the thickness of the catalyst (which is an unlikely scenario). The effective rates across the sample are averaged to produce the total rate of the reactor. This simulated total rate is fit against the maximum temperature of the reactor just as is done for the experimental data and a comparison of the resulting activation barrier plot is shown for the simulation and experiment in \cref{SI:Theory}a and b, respectively. As seen in \cref{SI:Theory}d and e, the trends in the purely photothermal rate qualitatively match the experimental value trends quite well. The temperature increases with increasing laser power, while the overall rate of the reaction decreases. At a given apparent surface temperature, the effective rate appears to decrease, and the activation barrier increases with higher energy photons and higher illumination power, albeit to a much lower magnitude. These results are based on a purely thermal model. There is no assumed change to the reaction mechanism. The apparent change in the activation barrier is only a manifestation of the fitting procedure inappropriately using a single temperature value to fit against a reaction rate measured from a complex temperature distribution. While there is qualitative agreement between these simulations and the experimental results, the observed increase in the apparent activation barrier is about 0.16 eV while the simulated increase is only 0.05 eV, a mismatch of a factor of 3. Two possible explanations for this quantitative mismatch are (a) the model used in the simulations is too simple to capture the complex photothermal response of the system or (b) there is also a photochemical effect changing reaction pathway and making the hydrogenation reaction more difficult to proceed. To point (a), it is certainly true that a more sophisticated model could be developed. In fact, the accuracy of IR camera measurements has been called into question, and this model could be made more reliable by computing an expected photothermal response of the system rather than relying on experimental observations.\cite{sivanAssistanceMetalNanoparticles2019,dubiThermalEffectsAlternative2020, unParametricStudyTemperature2020} The possibility of the experimentally observed maximum temperature simply causing an inaccuracy in the model was explored by allowing the simulated maximum temperature to vary outside of the observed value. Further, artificial increases in the simulated max temperature increased the modeled activation barrier increase, and a temperature increase of 200 °C above the baseline was needed to produce a simulated activation barrier equal to the experimentally observed activation barrier. As mentioned in the main text, the activation barrier was also found to vary with the spot size of the illumination with a tighter illumination profile showing a higher activation barrier. Curiously, the trend does not hold if the simulated reaction rate is plotted against the mean temperature in the Arrhenius fits instead, while an apparent increased activation barrier is still observed for the experimental data treated in such a manner. As for point (b), there have been a few reports of plasmonic photocatalysis increasing the activation barrier for hydrogenation reactions. Huang et al showed that the hydrogenation of styrene in solution on a core@shell AuNR@Pd structure was hindered under conditions favoring the production of hot electrons.\cite{huangUnravelingSurfacePlasmon2016} Yuan et al further showed that the simulated excited state pathway for NH3 dehydrogenation on a Pd surface in the presence of hot electrons was increased relative to the ground state barrier, supporting the concept that excited pathways could exhibit increased activation barriers.\cite{yuanPlasmonicPhotocatalysisChemically2022} In all, both of these explanations are plausible, but the apparent contribution of fitting inaccuracies resulting from in-plane temperature gradients in the catalyst bed is cause for concern when drawing conclusions about the effect of light on the observed apparent activation barrier of a reaction.

\subsection*{Rate Order Measurements}
\begin{figure}
\begin{center}
    \includegraphics[]{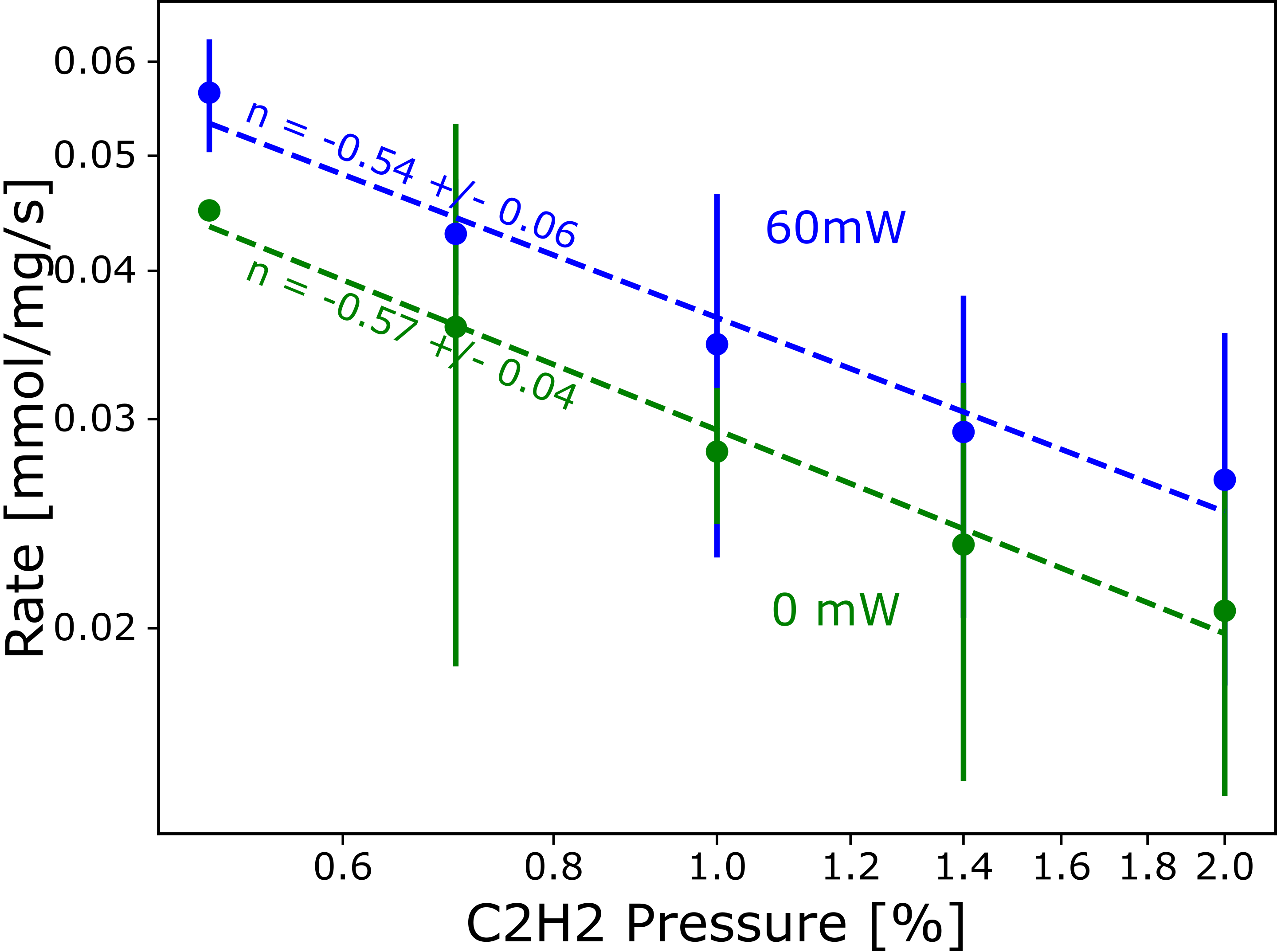}
    \caption{Effect of C$_2$H$_2$ of partial pressure on reaction rate. Rate order is represented as the slope next to fits. Both measurements are run at 380 K and the light on experiment uses 60 mW with a central wavelength of 530 nm with a bandwidth of 50 nm. The dark measurement is collected at a total flow of 50 sccm, while the illuminated sample is at 65 sccm to account for the additional activity.}
  \label{SI:C2H2_rate_order}
\end{center}
\end{figure}

While the apparent activation barrier measurements are one possible method for determining changes to the reaction mechanism of a photocatalytic reaction, rate order measurements may also elucidate mechanism changes. The rate order of a chemical reaction is the response of the system to a change in the concentration of one of the reactants. This response is fundamentally linked to the elementary steps of the reaction, hence, a change in the reaction mechanism may result in a change to the rate order. In this work, we explored the thermal and photo rate order for both the H$_2$ and C$_2$H$_2$ reactants. \cref{SI:C2H2_rate_order} shows the reaction rate as a function of C$_2$H$_2$ partial pressure. The system is heated to 380 K and the light on experiment additionally illuminates the system at 60 mW. The reaction rate can be written as $rate \propto [C_2H_2]^m * [H_2]^n$ in which m and n represent the rate order of the reaction with respect to C$_2$H$_2$ and H$_2$. The rate order is therefore determined by plotting the log of the rate as a function of the reactant pressure, fitting to a linear function, and computing the slope. As seen, the rate order for C$_2$H$_2$ in this case is around -0.5 which is inline with other experimental observations for thermal catalysis.\cite{zhangSelectiveHydrogenationAcetylene2014, adurizActivitySelectivityPd1990, luneauAchievingHighSelectivity2019} Upon illumination, the system performs nearly identically and the rate order does not change in a meaningful way, suggesting that plasmonic excitation has no impact on the role of C$_2$H$_2$ in the reaction.

\begin{figure}
\begin{center}
    \includegraphics[]{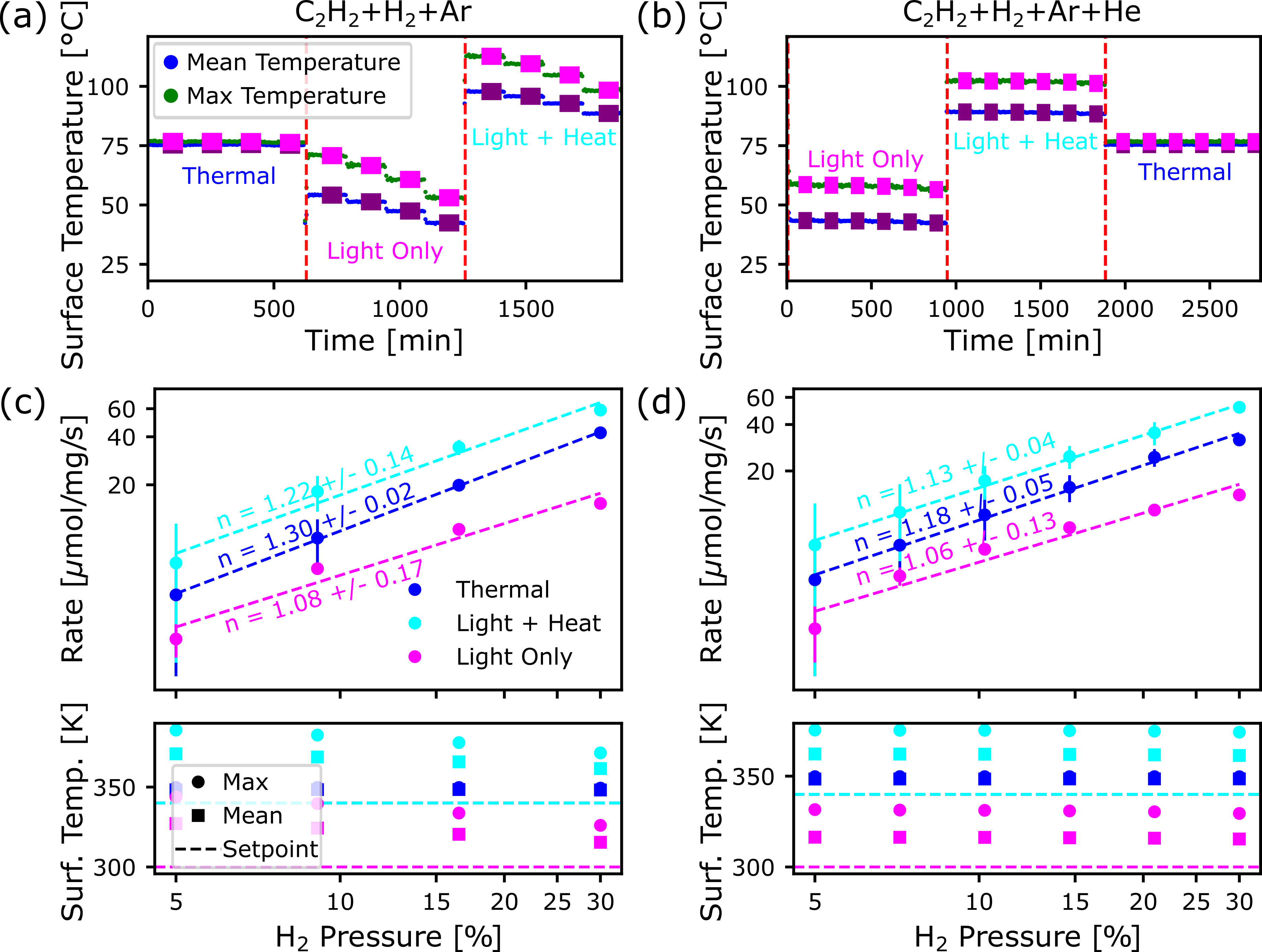}
    \caption{Effect of C$_2$H$_2$ of partial pressure on reaction rate. (a) temperature as a function of time over the course of the experiment. (b) the reaction rate as a function of the log of H$_2$ partial pressure as well as the temperature measured as each point. The max, mean, and global setpoint temperature is depicted by circles, squares, and dashed lines, respectively. (c) and (d) depicts similar data when He is added to the reactant stream to keep a consistent heat capacity of the mixture.}
  \label{SI:H2_rate_order}
\end{center}
\end{figure}

Swearer et al hypothesized increased selectivity derived from a reduction in H surface coverage because of plasmon induced desorption.\cite{swearerHeterometallicAntennaReactor2016} Furthermore, light-coupled ETEM experiments have shown plasmon driven H$_2$ desorption, suggesting that plasmons may in fact increase the rate of H$_2$ removal form the surface of Pd.\cite{vadaiInsituObservationPlasmoncontrolled2018, sytwuDrivingEnergeticallyUnfavorable2021} \cref{SI:H2_rate_order} demonstrates a similar rate order experiment performed by varying the H$_2$ content of the reactant stream. Parts (a) and (b) depict the initial experiments. The rate order with respect to H$_2$ is seen to decrease as the reaction becomes more purely optically driven. The thermal baseline shows a rate order slightly above 1, which is also consistent with other measurements made in the community on Pd based catalysts. Initially, these results seemed to support the above mentioned hypotheses, but careful attention needs to be paid to behavior of the system as a whole to effectively understand the validity of this interpretation. \cref{SI:H2_rate_order}a shows the measured mean and max surface temperature of the catalyst via IR imaging. While the purely thermal portion of the experiment shows a consistent and uniform temperature, the illuminated portions of the experiment show a deviation between the max and mean temperatures as well as variation over time. This variation is correlated directly to when the H$_2$ concentration changes over the course of the experiment. The thermal conductivity and specific heat of H$_2$ gas is notably high, and varying the amount of H$_2$ flowing into the system puts a changing thermal load on the system. Whereas the resistive heater used in the thermal portion is adjusted to account for the variation in H$_2$ flow, the laser provides a fixed power which contributes a fixed photothermal heat source to the system. Increased H$_2$ flow more effectively cools the system. To account for this discrepancy, He is added to the reactant mixture. He has a similar heat capacity to H$_2$. The effective heat capacity of the reactant mixture is estimated as a linear combination of the heat capacity of the 4 inputs streams, and the effective heat capacity is algorithmically kept at a constant value given the desired variation in H$_2$ content. As seen in \cref{SI:H2_rate_order}c \& d, the surface temperature is consistent throughout the optically illuminated portions of the experiment after the addition of He gas to the reactants. As a result, there is little variation in the observed rate order of the experiment. The change in the rate order seemed to be fully caused by the change in temperature. With increased H$_2$ concentration, the reaction rate was pushed down due to the decreased temperature of the catalyst. The rate order formula assumes a constant temperature, but this temperature change actually has an exponential impact on the reaction rate observed.

\subsection*{Chemical Selectivity}

\begin{figure}
\begin{center}
    \includegraphics[]{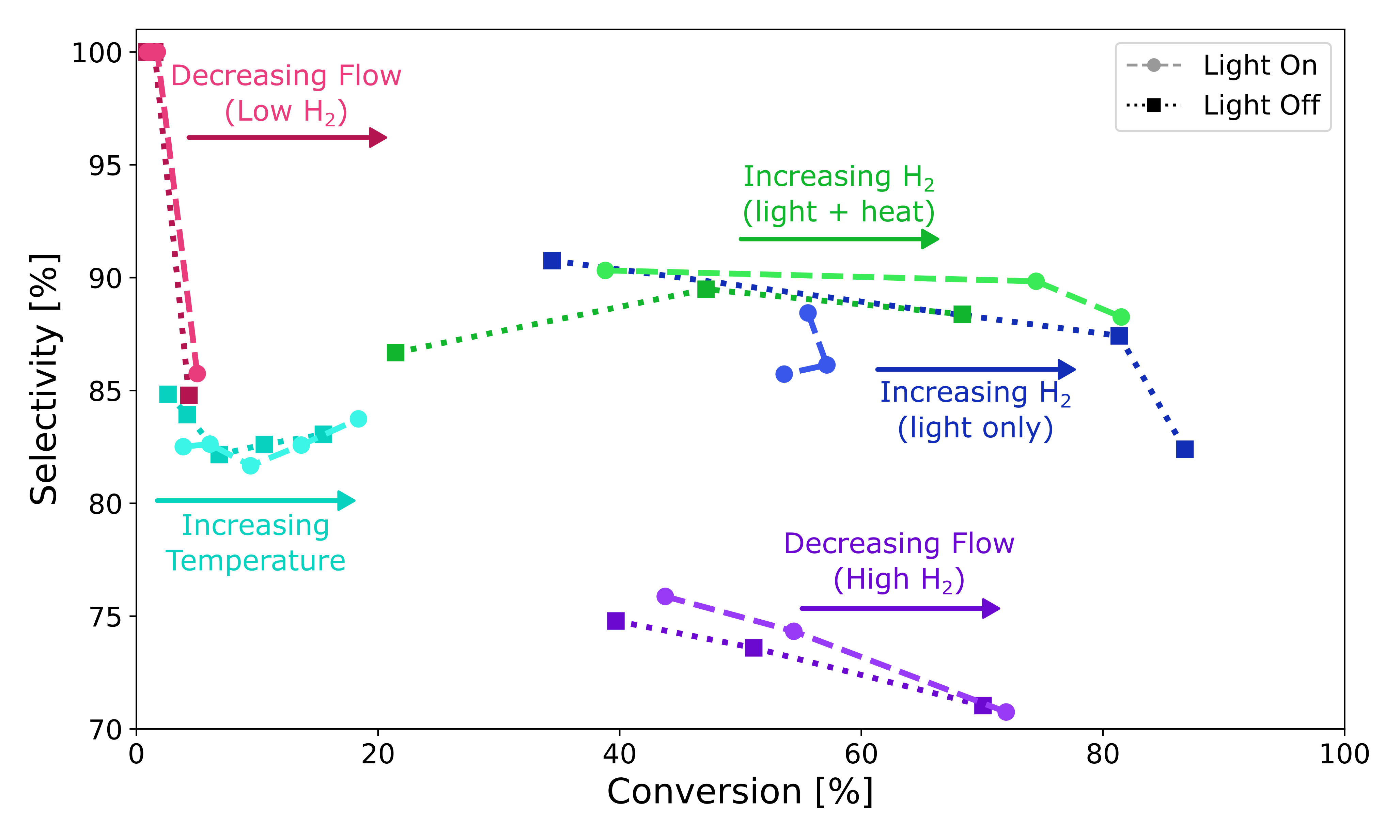}
    \caption{Chemical selectivity of C$_2$H$_2$ hydrogenation under various conditions as a function of conversion. Selectivity is calculated as the portion of C$_2$H$_4$ divided by the sum of C$_2$H$_4$ and C$_2$H$_6$. No appreciable amount of other byproducts is measured.}
  \label{SI:X_vs_S}
\end{center}
\end{figure}

Perhaps the most compelling reported effect of plasmon driven hydrogenation chemistry is an increase in the chemical selectivity of the reaction.\cite{swearerHeterometallicAntennaReactor2016, peirisControllingSelectivityPlasmonic2023} C$_2$H$_2$ hydrogenation operates in two distinct regimes. When the H$_2$ concentration is kept relatively high the reaction primarily produces C$_2$H$_6$ as an undesirable byproduct.\cite{McCue2015} This is caused by the over hydrogenation of the C$_2$H$_2$ molecule, in which the desired C$_2$H$_4$ intermediate undergoes further hydrogenation prior to desorption. This reaction pathway is particularly pronounced at high conversion. Typically, the binding energy of C$_2$H$_2$ is higher than that of C$_2$H$_4$, displacing it before additional hydrogenation steps can occur. When the C$_2$H$_2$ content has been sufficiently depleted, this preferential bonding no longer encouraged chemical selectivity. As such, it is critical to evaluate chemical selectivity at the same conversion values to properly determine the selectivity of the catalyst. Additionally, there are a number of ways in which the conversion of the reaction can be adjusted. \cref{SI:X_vs_S} depicts the chemical selectivity of the AuPd catalyst under a variety of conditions as a function of conversion. Each color represents a different set of conditions; the square markers indicate thermal reactions while the circle markers indicate illuminated conditions. In all cases, it is clear that illumination does not promote higher chemical selectivity. Previous reports did show increased chemical selectivity over antenna/reactor style Pd based catalysts when converted into an analysis of this format.\cite{swearerHeterometallicAntennaReactor2016} It should be noted, however, that the conversion range for the light driven reaction (limited by laser power) is much lower in this case, and more thorough exploration of the system is desired to confirm the effect. A possible origin of this discrepancy is the alloyed nature of the catalyst explored here. One cause of an unselective catalyst of this nature is the existence of a hydride phase within the Pd component of the catalyst.\cite{teschnerRolesSubsurfaceCarbon2008, teschnerAlkyneHydrogenationPd2006, borodzinskiSelectiveHydrogenationEthyne2007, armbrusterHowControlSelectivity2012, tewRolesCarbideHydride2011} Very low Pd content systems have a lower affinity for H absorption and hinder the formation of a hydride phase inherently. Plasmon excitation is reported to expel this hydride phase, which has a lower chemical selectivity for C$_2$H$_4$. It is possible that initial reports of increased selectivity in these antenna reactor-based catalysts were actually observing light-driven phase transformation impacting the chemical selectivity of the catalysts. In comparison, such phase transformations would not be expected in our system. In fact, the H surface content of our system should be particularly low, as the Pd coverage is in the single atom regime, based on concentration. As such, the H coverage may already be low enough to encourage selective hydrogenation chemistry, and further H$_2$ desorption by plasmonic action would not improve performance. A more quantitative examination of the H surface coverage under these different conditions would better elucidate the origins of these selectivity enhancements.

\subsection*{Computer Interface}
\label{computer_interface}
The above described hardware components are operated exclusively through the Catalight programming interface, with the exception of the thermal camera. The thermal camera is set to collect at the beginning of the experiment using FLIR Studio and the data is imported during the analysis. Detailed descriptions of the computer connections to the hardware can be found in the Catalight documentation. In brief, the Alicat system and Harrick heater (via the built-in Watlow controller) are connected to using independently developed Python packages called "alicat" and "pywatlow". These packages are imported to Catalight. The GC is controlled through the PeakSimple software interface using the manufacterer provided .NET assembly. In practice, this involve importing the manufacturer provided .dll file, here using the "pythonnet" package. The relevant files are included in the Catalight distribution files and do not need to be independently downloaded by the user. Similarly, the NKT laser is controlled via a manufacturer provided .dll file and imported using a manufacturer provided python script. The simplify the user interface, we seperately developed a second python package called "nkt\_tools" which wraps over the manufacturer provided interface in a more user friendly manner.\cite{nkt_tools} This second python package also contains all relevant files needed and is imported to Catalight as a dependency.

\section*{Automation Script}
\label{script}

The below script was used to perform the apparent activation barrier measurement using the Catalight \textcolor{red}{v0.1.0} package.\cite{catalight_package}
\begin{lstlisting}[language=Python, caption = Necessary imports]
import os
import time

import numpy as np

from catalight.equipment.gas_control.alicat import Gas_System
from catalight.equipment.light_sources.nkt_system import NKT_System
from catalight.equipment.heating.watlow import Heater
from catalight.equipment.gc_control.sri_gc import GC_Connector
from catalight.equipment.experiment_control import Experiment
\end{lstlisting}

\begin{lstlisting}[language=Python, firstnumber=10, caption= Function definitions]
def initialize_equipment():
    ctrl_file = (r"C:\Users\dionn\GC\Control_Files\HayN_C$_2$H$_2$_Hydrogenation"
                 r"\20221106_C$_2$H$_2$_Hydro_HayN_TCD_off.CON")
    gc_connector = GC_Connector(ctrl_file)
    laser_controller = NKT_System()
    gas_controller = Gas_System()
    heater = Heater()
    return (gc_connector, laser_controller, gas_controller, heater)


def shut_down(eqpt_list):
    print('Shutting Down Equipment')
    gc_connector, laser_controller, gas_controller, heater = eqpt_list
    laser_controller.shut_down()
    heater.shut_down()
    gas_controller.shut_down()


def run_study(expt_list, eqpt_list):
    for expt in expt_list:
        try:
            expt.run_experiment()
        except Exception as e:
            shut_down(eqpt_list)
            raise (e)
\end{lstlisting}

\begin{lstlisting}[language=Python, firstnumber=35, caption=Main Script]
if __name__ == "__main__":
    eqpt_list = initialize_equipment()  # Initialize equipment
    # Create folder to save data into
    sample_name = '20230504_Au95Pd5_4wt_3mg'
    main_fol = os.path.join(r"C:\Users\dionn\GC\GC_Data\20231009", sample_name)
    os.makedirs(main_fol, exist_ok=True)

    # Run a pre-experiment reduction
    reduction = Experiment(eqpt_list)
    reduction.expt_type = 'stability_test'
    reduction.gas_type = ['Ar', 'C$_2$H$_2$', 'H$_2$', 'Ar']
    reduction.temp = [300+273]
    reduction.gas_comp = [[0, 0, 0.05, 0.95]]
    reduction.tot_flow = [50]
    reduction.sample_rate = 30
    reduction.sample_set_size = 4
    reduction.t_steady_state = 30
    reduction.sample_name = sample_name
    reduction.create_dirs(main_fol)
    try:
        reduction.run_experiment()
    except Exception as e:
        shut_down(eqpt_list)
        raise (e)

    # Main experiment
    for wavelength in range(480, 680+1, 50):
        for power in range(0, 60+1, 20):
            expt = Experiment(eqpt_list)
            expt.expt_type = 'temp_sweep'
            expt.temp = list(np.arange(340, 380+1, 10))
            expt.wavelength = [wavelength]
            expt.power = [power]
            expt.bandwidth = [50]
            expt.gas_type = ['Ar', 'C$_2$H$_2$', 'H$_2$', 'N2']
            expt.gas_comp = [[1-0.06, 0.01, 0.05, 0]]
            expt.tot_flow = [10]
            expt.sample_set_size = 4
            expt.t_steady_state = 30
            expt.sample_name = sample_name
            expt.create_dirs(main_fol)
            try:
                expt.run_experiment()
            except Exception as e:
                shut_down(eqpt_list)
                raise (e)
    shut_down(eqpt_list)
\end{lstlisting}

\setlength{\bibsep}{0.0cm}
\bibliographystyle{Wiley-chemistry}
\bibliography{references}